\documentclass[aps,prd,amsmath,twocolumn,longbibliography]{revtex4-1}

\usepackage{graphicx}
\usepackage{amssymb,color}
\renewcommand{\bar}[1]{\overline{#1}}
\renewcommand{\bar}[1]{\overline{#1}}
\newcommand{\half}{{\frac{1}{2}}}

\def\e{\epsilon}
\def\Dslash{\raise.15ex\hbox{/}\kern-.7em D}
\def\Pslash{\raise.15ex\hbox{/}\kern-.7em P}

  \def\Cv{{\cal V}}

\def\s{\sigma}

\def\t{\tau}
\newcommand{\la}{\langle}
\newcommand{\ra}{\rangle}
\newcommand{\bfr}{{\bf r}}

\newcommand{\ben}{\begin{displaymath}}
\newcommand{\een}{\end{displaymath}}
\newcommand{\be}{\begin{equation}}
\newcommand{\ee}{\end{equation}}
\newcommand{\bea}{\begin{eqnarray}}
\newcommand{\eea}{\end{eqnarray}}

\newcommand{\eq}[1]{Eq.~(\ref{#1})}

\newcommand{\bfp}{{\bf p}}
\newcommand{\bfP}{{\bf P}}
\newcommand{\bfq}{{\bf q}}            \def\bfk{{\bf k}}

\newcommand {\bfnabla}{\mbox{\boldmath$\nabla$}}

\def\g{\gamma}\def\r{\rho}\def\L{\Lambda}\def\a{\alpha}\def\l{\lambda}

\def\m{\mu}

\newcommand{\beqn}{\begin{equation}}
\newcommand{\eeqn}{\end{equation}}

 \newcommand{\vlowk}{V_{{\rm low}\,k}}
\def\CO{{\cal O}_{\rm DIS}}

\newcommand{\fmi}{\, \text{fm}^{-1}}

\def\Pm{P_{\rm miss}}\def\n{\nu}
 \newcommand{\Oop}{{\cal O}}\def\D{\Delta}
 \def\bfQ{{\bf Q}}

\begin{document}

\title{\bf  \hskip10cm NT@UW-20-06\\
Discovery {\it vs.}  Precision in  Nuclear Physics- A Tale of Three Scales}

\author{Gerald A. Miller}

\affiliation{ 
Department of Physics,
University of Washington, Seattle, WA 98195-1560, USA}

\date{\today}

\begin{abstract}
At least   three length scales are important in  gaining    a complete understanding of the physics of nuclei. These are the radius of the nucleus, the average inter-nucleon separation   distance, and the size of the nucleon. The connections between the different scales are examined by using examples that demonstrate  the direct connection between short-distance and high momentum transfer physics and also that significant high momentum content of wave functions is inevitable.  The nuclear size is connected via the independent-pair approximation to the nucleon-nucleon separation distance, and this distance is  connected via the concept of virtuality to the EMC effect. An explanation of the latter is presented in terms of light-front holographic wave functions of QCD.  The net result is that  the three scales are closely related, so that a narrow focus on any given specific range of scales may prevent an understanding of the fundamental  origins of nuclear properties. 
It is also determined that, under certain suitable conditions, experiments are able to measure the momentum dependence of wave functions.
 \end{abstract}  
\maketitle     
\noindent

\section{Introduction}
 In studying atomic nuclei one encounters three different length scales: the nuclear radius $R_A (\approx 5$ fm for a heavy nucleus),  the average separation between nucleons at the centers of nuclei $d\approx 1.7$  fm, and  the nucleon radius, $r_N\approx 0.84 $ fm. The pion Compton wave length, $1/m_\pi=$ 1.4 fm is close to  $d$, so is not a separate scale.  The correlation length associated with the Fermi momentum, $\approx\pi/k_F$~\cite{Bohr}  is also of the order of $d$. 
 \\

 
 The general  modern trend of theorists is to focus on each length scale of a given subject using the techniques of effective field theory. The main idea 
 (see {\it e.g.}~\cite{Georgi:1994qn}) is:  
  if there are parameters that are very large or very small compared to the physical quantities (with the same dimension) of interest,   one  may get a simpler approximate description  of the physics by setting the small parameters to zero and the large parameters to infinity. Then the finite effects of the large parameters can be included as small perturbations about the simple approximate starting point.\\

This  scale separation is a common technique (see {\it e.g.}  \cite{Cohen:2019wxr})  in which physics at large distances  is assumed not to  depend  on physics at shorter distances.   A famous example is the weak interaction in which the effects of $W$ and $Z$ boson exchanges can be treated as contact (zero-ranged) interactions at low energies.  The general philosophy is  that if one is working at a  low mass scale $m$ one doesn't need to consider dynamics at a mass scale $\L \gg m $. Or in terms of distances: the long distance scale
$1/m$ must be very much greater than the short distance scale $1/\L $. In other words, there must be a large separation of scales for effective field theory techniques to be maximally efficient. 
In nuclear physics  the scale separation is not very large--the values of  relevant distances are not widely separated.  \\

In using  effective field theory, theorists concentrate  on a given range of length scales. A typical procedure is  to make robust calculations  that enable firm predictions. These are then tested by experiments, and the results may confirm the theories or (more likely) lead to revision of the theories. Another scenario, in which experiment leads,  is that an experiment discovers an unexpected phenomenon, such as the Rutherford's discovery of the atomic nucleus  or the SLAC-MIT discovery of quarks within the nucleon~\cite{Bloom:1969kc,Friedman:1972sy}. \\

The two approaches of the previous paragraph can be summarized as  precision {\it vs} discovery. The effective field theory approach of working within a given scale is aptly suited for precision work. In contrast,  discovery of new phenomena is not well treated by scale separation techniques because  new phenomena are often related to discovering  a new relevant scale.\\

I comment  on the precision approach.
Much current activity  in  precision nuclear structure calculations is based on using low energy, long length scale treatments. These began with interactions, known as $\vlowk$, that  use 
renormalization group  transformations that lower a cutoff 
in relative momentum    to derive NN potentials
with vanishing matrix elements for momenta above the cutoff.
Such interactions 
show greatly enhanced convergence properties in nuclear few- and 
many-body systems for cutoffs of order $\Lambda = 2\fmi$ or
lower~\cite{Bogner:2001gq,Bogner:2003wn,Bogner:2001jn,Nogga:2004ab,Bogner:2005sn}.
 Later calculations use renormalization group methods  to soften interactions
in nuclear systems. This  extends the range of many
computational methods and qualitatively improves their convergence
patterns~\cite{Bogner:2009bt}.
The similarity renormalization group (SRG) \cite{Glazek:1993rc,Szpigel:2000xj,Bogner:2006pc}
does this by systematically evolving 
Hamiltonians via a continuous series of unitary transformations 
chosen to decouple the
high- and low-energy matrix elements of a given interaction
\cite{Jurgenson:2007td,Jurgenson:2009qs}.\\

However, many  conventional NN
potentials,  feature strong short-range repulsion~\cite{RevModPhys.81.1773}.
This is supported by some lattice gauge QCD calculations
~\cite{PhysRevLett.99.022001,Aoki:2009ji,Murano:2011nz,Doi:2015oha,PhysRevD.98.038501,Aoki:2017byw}.  
The repulsion causes  bound states with very low energies 
(such as the deuteron) to have important contributions to the binding
and other properties from high-momentum components. \\

In Ref.~\cite{Benhar:1993ja}, the authors calculate cross sections
for electron scattering from light nuclei.    They conclude:
``and thus the data confirm the existence of high-momentum components
in the deuteron wave function''.  The high-momentum components of the deuteron lead to inclusive electron-scattering  cross section ratios with simple scaling 
properties~\cite{PhysRevC.48.2451}.  That reference finds significant ``evidence for the dominance of short-range correlations in nuclei". 
Ref.~\cite{Bogner:2007jb} argued that the statement of   Ref.~\cite{Benhar:1993ja} (and by implication that of Ref.~\cite{PhysRevC.48.2451})  is not correct because  wave functions are not observables.
Similarly Ref.~\cite{Furnstahl:2001xq} argued that nuclear momentum distributions are not observable.
It is certainly true that wave functions are not observable quantities, but cross sections are observables.\\

There are prominent examples that momentum-space wave functions are  closely related  to cross sections. Showing that the cross section of the photo-electric effect in hydrogen  is proportional to the square of the momentum-space ground-state wave function of hydrogen is a text-book problem~\cite{Sakurai:2011zz,Gottfried}.
The modern version of the photo-electric effect is called Angle Resolved Photoemission Spectroscopy (ARPES) a technique 
that is well-known, see {\it e.g.} Ref.~\cite{RevModPhys.75.473}, to yield information of about the momentum and energy states of electrons in materials.
The statement that  measurements of cross sections can be used to learn about wave functions violates no principles of quantum mechanics.
\\

One of the purposes of this paper is to exemplify how the use of the impulse approximation simplifies the connection between cross sections and wave functions for 
nuclear processes at  high momentum transfer.
If the kinematics are correctly chosen the effects of various processes that are not directly related to wave functions can be minimized~\cite{Schmidt:2020kcl}, so that in effect 
measuring cross section measures important properties of wave functions.   See Sects.~IV, VI, and VII.\\
 

The principle concern of the present epistle is that 
current 
 experiments involving nuclei cover all the  three scales mentioned above. Deep inelastic scattering experiments on nuclei, involving  squares of four momentum transfers ($Q^2$) between 10 and hundreds of GeV$^2$  have shown that the quark properties (quark distributions) of nucleons bound in nuclei are different than those of free nucleons. This phenomenon is known as the  EMC effect; see {\it e.g.} the review~\cite{ Hen:2016kwk}. The effect is not large, of order 10-15\%, but is of fundamental interest because it involves the influence of nuclear properties on scales that resolve the nucleon size.   
 But  scales larger than the nucleon size are relevant because modifications of nucleon structure must be caused by interactions with nearby nucleons. Indeed, after the nucleon size, the next largest length is the inter-nucleon separation length, $d$. This is the scale associated with short range correlations between nucleons. Therefore the EMC effect is naturally connected with short range correlations between nucleons. But the inter-nucleon separation is not very much smaller than that of the nuclear size. This means that  effects involving the entire nucleus cannot be disregarded. Such effects are known as mean-field effects in which each nucleon moves in the mean field provided by other nucleons. Understanding the EMC effect involves understanding physics at all three length scales.\\

Here is an 
outline of the remainder  of this paper. Sect.~II presents a short review of the modern technique of softening the nucleon-nucleon interactions  to simplify calculations of low-energy nuclear properties. The consequence of this softening is  the  hardening of the leptonic  interactions that probe the system. Sect.~III is concerned with the largest of the three nuclear  distance scales--the nuclear radius. This is followed by a discussion of the physics of the separation between  two nucleons in bound states,  Sec.~IV.
The consequent nuclear manifestations are  discussed in Sect.~V. This involves understanding the connection between the physics of short distances and high momentum. It is shown that the momentum dependence of wave functions can in principle be observed by measuring elastic form factors. Next, Sect.~VI discusses the $(e,e'p)$ reaction as a discovery mechanism for the physics of the two-nucleon separation distance. The concept of virtuality (the difference between the square of the four-momentum and the square of the mass)  as a connection between the scale of the
 two-nucleon separation-distance and the nucleon size is introduced in Sec.~VII.  The connection between virtuality and the EMC effect is elucidated in Sect.~VIII. Finally, a summary is presented, Sec.~IX.\\

I aim  to  explain the basic ideas as clearly as possible by using simple examples. There is no intent to present  detailed state-of-the-art calculations. A separate  direction, not discussed here,    is that precision nuclear structure calculations  can be used in the aid of discovery, such as in the searches for neutrinoless double beta decay~\cite{RevModPhys.80.481}  and/or beyond the standard model particles  \cite{Kozaczuk:2016nma}.


\section{Softened NN Potentials and Hardened Interaction Operators}
\label{sect:EdepRG}

The use of scale separation began with applying chiral effective field theory 
to the nucleon-nucleon interaction~\cite{Ordonez:1992xp,Ordonez:1993tn,Ordonez:1995rz}.
This work stimulated many efforts, see  {\it e.g.} the reviews~\cite{Bedaque:2002mn,RevModPhys.92.025004}. \\

 Another approach is to use low momentum nucleon-nucleon 
 interactions~\cite{Bogner:2001gq,Bogner:2001jn,Bogner:2003wn,Bogner:2005sn,Bogner:2006vp,Bogner:2007jb,Bogner:2009bt}.
  After that came the
 similarity renormalization group \cite{Glazek:1993rc,Szpigel:2000xj,Bogner:2006pc,Jurgenson:2007td,Jurgenson:2009qs,Bogner:2009bt,Anderson:2010aq}
  which involves a  unitary transformation on nucleon-nucleon interactions and the operators that represent observable quantities.   The present  section is intended as a brief review of the latter  two techniques, with emphasis placed on the necessary transformations of the operators that probe the system.
  \\


Let's begin  by describing a simple cutoff theory as described  by  Bogner {\it et al.}~\cite{Bogner:2001gq} who found that  the effective interactions constructed from various high precision nucleon-nucleon interaction models  are identical. 
Their approach is to obtain the half-off shell $T$-matrix via the equation
\bea &T(k',k;k^2)=V_{\rm low\,k}(k',k) \nonumber\\&+
{2\over \pi}
{\cal P}
\int_0^\L
{V_{\rm\,low\,k}(k',p)T(p,k;k^2)\over k^2-p^2}
p^2dp\label{vlk} \eea
for a single partial wave in which $k'$ and $k$ denote the relative momenta of the outgoing and incoming nucleons, and the mass of the nucleon is taken to be unity. Furthermore, {\it all} momenta are constrained to lie below the cutoff $\L$.  A specific formalism was developed to obtain $V_{\rm low\,k}$ from the initial bare interaction $V$. This construction enforces the condition that the half-off-shell $T$-matrix is independent of the cutoff parameter $\L$.\\

As a consequence 
of the cutoff independence of the  half-off-shell  $T$-matrix, the interacting scattering eigenstates of the
 low-momentum Hamiltonian $H^\L \equiv H_0 + V_{\rm low\,k}$ (where $H_0$ is the kinetic energy operator)
 are equal to the
low-momentum projections of the corresponding scattering and bound eigenstates, $|\Psi_k\ra,\,|\Psi_B\ra$  of the original  Hamiltonian, $H_0+V$ \cite{Bogner:2008xh}. This means that
$|\chi_k\rangle  = P|\Psi_k\rangle $ , with an analogous relation for bound states,
  \bea|\chi^\L_B\ra = P|\Psi_B\ra,\label{proj}\eea
  where $P$ is an projection operator onto states of relative momenta less than $\L$.  The consequences of the projection operator $P$ in \eq{proj} are  studied below.\\

Suppose the  system is probed by an interaction operator, here defined as $\Oop$. The procedure invoked by using \eq{vlk}  leads to the requirement that $\Oop$    is to be  dressed. The  transformation corresponding to the first in the series of three transformations used to derive  a $V_{\rm low\,k}$ that is Hermitian and independent of energy~\cite{Bogner:2001jn}  is:
\bea \Oop\to (1 +H_{PQ}{1\over E-H_{QQ}})\Oop (1+ {1\over E -H_{QQ}}H_{QP}) \label{renorm}
,\eea
where $Q=I-P$ and $H_{QQ}=QHQ$, {\it etc.}   This projection operator  procedure maintains the correct value of the matrix elements of $\Oop$, and is sufficient for present explicative purposes.\\

The key feature of \eq{renorm} is 
that the effects of any high momentum component ($Q$-space) in the wave function that are removed by using \eq{proj} as the wave function are incorporated in the probe operator. Thus, the  probe operator must  be hardened by the softening of the two-nucleon potential.
\\

The use of   $V_{\rm low\,k}$ to soften the NN potential  was followed by  
renormalization group methods~\cite{Bogner:2009bt}.
The similarity renormalization group (SRG) \cite{Glazek:1993rc,Szpigel:2000xj,Glazek:2001uw,Bogner:2006pc}
achieves softening  by  evolving 
Hamiltonians with a continuous series of unitary transformations 
chosen to decouple the
high- and low-energy matrix elements of a given interaction
\cite{Jurgenson:2007td,Jurgenson:2009qs}.
Thus 
\beqn
  H_{s}=U_{s}H U^{\dagger}_{s}=H_0+V_{s}
  \;,
  \label{first}
\eeqn
with $H=H_0+V\equiv H_{s=0}$, and $H_0$ is the kinetic energy operator. The generator of the transformation is $\eta_s={dU_s\over ds}U_s^\dagger=-\eta_s^\dagger$
and  ${dH_s\over ds}=[\eta_s,H_s]$,
 The choice of the anti-Hermitian operator $\eta_s $ 
as  $\eta_s=[H_0,V_s]$ has proved to be convenient and  is used here. The kinetic energy operator is  not changed by the transformation.\\

Ref.~\cite{Anderson:2010aq} correctly emphasized that 
when using  the wave functions produced by 
SRG-evolved interactions to calculate other matrix elements of
interest, the associated unitary transformation of operators must be implemented. See also~\cite{Tropiano:2020zwb}.
The evolution of any operator $\Oop\equiv\Oop_{s=0}$ is given by the same
unitary transformation used to evolve the Hamiltonian~\cite{Szpigel:2000xj,Bogner:2007jb},
\beqn
  \Oop_{s}=U_{s}\Oop_{s=0}U^{\dagger}_{s}
  \;,
   \label{Otransform}
\eeqn
which  obeys 
the general operator SRG equation
\beqn
\frac{d\Oop_{s}}{ds}=[[H_0,V_s],\Oop_{s}]
  \;.
  \label{flow}
\eeqn
 If implemented without approximation, unitary transformations
preserve matrix elements of the operators that define observables.\\ 

  
The focus here is on   the calculation of observables. Consider an operator $\Oop$, consistent with the bare Hamiltonian $H=H_0+V,$ that probes the system.
The applications discussed here involve the interactions between a lepton probe and the system.
The    operator flow equation, \eq{flow}, is rewritten using the Jacobi identity as
\beqn
  \frac{d\Oop_{s}}{ds}   
   =[H_0,[V_s,\Oop_{s}]]+[V_s,[\Oop_s,H_0]]\label{J}
 ,
\eeqn
with the boundary condition $\Oop_{s=0}=\Oop$. 
 To illustrate the main idea, let's take  $ \Oop$ to depend only on coordinate-space operators, and the bare potential to be local. Then for $s=0,\,  [V,\Oop]=0$, and  for a   system in its center of mass
\bea &[\Oop,H_0]={1\over 2M_r}(\nabla^2 \Oop+2 \bfnabla \Oop\cdot\bfnabla)\\&
[V,  [\Oop,H_0]]={-1\over M_r} 
\bfnabla V\cdot\bfnabla\Oop
\eea
with $M_r$ the reduced nucleon mass.
To first-order in $s$
\bea \Oop_s=\Oop -{s M_r}\bfnabla V\cdot\bfnabla\Oop,\label{fo}\eea
and one sees immediately  that the evolution converts a one-body operator to a two-body operator. The factor of $M_r$ arises from converting the units here to those of 
\cite{Anderson:2010aq} in which $s=0.2 $ fm$^4$. A term of first-order in $s$ that arises from the $s-$dependence of the potential vanishes here,   as shown in the Appendix,   \\

To see the explicit  effect of hardening of the interaction operator, let $\Oop$ be the momentum transfer operator $e^{i\l\bfq\cdot\bfr}$, (in which the real-valued  parameter $\l$ accounts for using the relative coordinate) then $\Oop_s$ acquires a factor of $\bfq$ which gets larger as the momentum transfer increases. \\

For an $A$-nucleon system this evolution  procedure
would turn  a one-body operator into an $A$ body operator, as explained in Ref.~\cite{Anderson:2010aq}.\\

The stage is now set for the discussion of lepton-nucleus  scattering in terms of the three scales of nuclear physics, starting with the largest and proceeding to the smallest.

\section{Discovery of Non-Zero Nuclear Sizes}

This Section is concerned with the largest of the three nuclear scales- the nuclear radius. Though small on the scale of atomic sizes, the nuclear radius is large in the present context.\\

Hofstadter, as part of his  Nobel-prize winning work, showed \cite{Hofstadter:1956qs,Hofstadter:1957wk}  (in first Born approximation) that 
the electron-nucleus scattering cross section $\s_s(\theta)$ was proportional to the square of the three-dimensional Fourier transform
of the nuclear charge density:
\bea  \s_s(\theta)\propto \left|\int d^3r \r(r) e^{\i\bfq\cdot\bfr}\right|^2,\label{Born}\eea
where $\r(r)$ is the nuclear charge density as a function of the separation from the center of the nucleus. Relativistic corrections are small for nuclear targets \cite{Miller:2009sg}. The three-dimensional  integral appearing  in \eq{Born} is defined to be the form factor $F(q)$. Electron scattering, in measuring the difference between the  form factor and unity, showed that the nucleus was not a point charge, as it would have been in a lowest-order effective field theory treatment. Importantly, electron scattering   was one of the main methods to determine the spatial extent of nuclear charge distributions~\cite{article}.
\\
 
For large nuclei the density is well-approximated by a Woods-Saxon (Fermi) form $\r(r)={\r_0\over 1+e^{(r-R)/a}}$. For nuclei wth $A>20$, $\r_0=0.17 {Z\over A}
$fm$^{-3},\,r=1.1\,{\rm fm}  A^{1/3}$  and $a=0.54 $ fm~\cite{article}. 
The nuclear diffuseness   $a$   can be understood as follows.  Each nuclear single-particle state falls exponentially with distance away from the nuclear center. Thus  the density falls a $e^{-r/a}$ for large $r$, with $a\approx 1/2/\sqrt{2M B}$ with $B$  the average binding energy at the center of the nucleus $B=16 $ MeV and $M$ the nucleon mass, $a=0.57 $ fm, which is close to empirical values and close to the size of the nucleon. The distance scale could instead be taken as the surface thickness, $t=4.4 \,a\approx 2 $ fm, the distance over which the density drops for 90 to 10 \% of its maximum value. The value of $t$ is close to the nucleon-nucleon separation distance.
Thus the two smallest nuclear size scales   enters in understanding the largest nuclear radius. This is an example of the principle that all of three nuclear distance scales are connected on a deep level.\\

The remainder of this Section is concerned with understanding the role of $a$, and in examining the effects of softening the nucleon-nucleon interaction. 

\subsection{Effects of the Diffuseness}
Examining the  effects of $a$  is simplified by using the nuclear shape as
  parameterized by  the symmetrized   Fermi form~\cite{PhysRevC.36.1105}:
\bea &\r(r)
=\r_0\frac{ \sinh \left(\frac{c}{a}\right)}{
   \left(\cosh \left(\frac{c}{a}\right)+\cosh \left(\frac{r}{a}\right)\right)}\\&
\r_0=\frac{3}{4 \pi  c^3 \left(\frac{\pi ^2 a^2}{c^2}+1\right)},\label{tpf}
\eea
which,  for large nuclei with $c/a\gg1$, is indistinguishable from the usual Fermi form.
The Fourier transform of this function yields 
the nuclear form factor  given by 
\bea&
F(q)=\r_0 {4\pi^2a c \over q \sinh{\pi a q} } (\pi  a/c \coth (\pi  a q) \sin (c
   q)-\cos (c q) ).\nonumber\\&\eea
 The mean-square radius defined by 
\bea \la r^2\ra\equiv \int d^3 r\r(r)r^2={1\over5}(3 c^2+7 a^2 \pi^2).\eea

Using  $c=6.38 $ fm and $a=0.535 $ fm for the Gold nucleus \cite{PhysRev.101.1131} as an example,
we see that
$ \la r^2\ra = 28.4 \,{\rm fm}^2$ with the term proportional to $a^2$ contributing  about  $4$ fm$^2$.  Thus the small scale of $a$ contributes about 14\% to the mean square radius and about 7\% to the rms radius. The small distance scale is important. Another example of importance is that 
the diffuseness $a$ leads to an exponential fall-off with $q$:
\bea \lim_{q\to\infty}F(q)={e^{-\pi a q}\cos cq \over q}.\eea
.

\subsection{Influence of the Softened Nucleon-Nucleon Interaction}
Let's examine the effect of the unitary transformation on the nuclear form factor. Use \eq{fo}  with the probe operator $\Oop=e^{i \bfq\cdot \bfr}$, taking $(A-1)/A\to1$,
where $\bfr$ represents the nucleon position operator and $\bfq$ is the momentum transfer. Evaluating the matrix element of the softened nucleon-nucleon potential  operator  in the  nuclear ground states leads, via the Hartree-Fock approximation,  to a nucleon-nucleus, shell-model interaction which is taken as a local potential,   $U(r)$. Such a mean-field  potential has the shape of the nuclear density, {\it e.g.} 
\eq{tpf}, with a central depth of about  57 MeV~\cite{Krane:1987ky}. Non-locaility of the mean field is neglected here to simplify the presentation.
\\

One finds  from \eq{fo} that
\bea\Oop \approx  (1- i \bfq\cdot \hat \bfr s M  { U'})\,e^{i\bfq\cdot\bfr}.\label{foA}\eea 
This first-order change in $\Oop$ is accompanied by a first-order change in the wave function, so that in principle the computed form factor is not modified by
the unitary transformation.  \\

The purpose here is {\it only} to illustrate the effect of the hardening of the interaction  
 caused by transformations such as those of \eq{fo}. 
 Therefore I compute the change in the form factor, $\D F$ caused by including the  
 second term of \eq{foA}. This change is given by 
\bea &\D F(q)=- {8\pi (s M)\over3} q \int r^2 dr \r(r) {dU\over dr}j_1(qr),\label{foa1}\eea
with value of  $s=0.2$ fm$^4$~\cite{Anderson:2010aq}.
A comparison between $F(q)$ and $F(q)+\D F(q)$ is made in Fig.~1.  The term $\Delta F$ is negligible for $q<1 $ fm$^{-1}$, but is about a 10\% effect for 
1.3 $\fmi$ and dominates for   $q>2\fmi$.     %
If $\D F$ is large compared with $F$ it is necessary to compute higher order terms, so the  details would change. Nevertheless, Fig.~1
 demonstrates the hardening of the probe interaction that occurs for large  values of the momentum transfer.\\



  \begin{figure}[h]\label{Nuclear}
\includegraphics[width=5.99cm,height=3.9253cm]{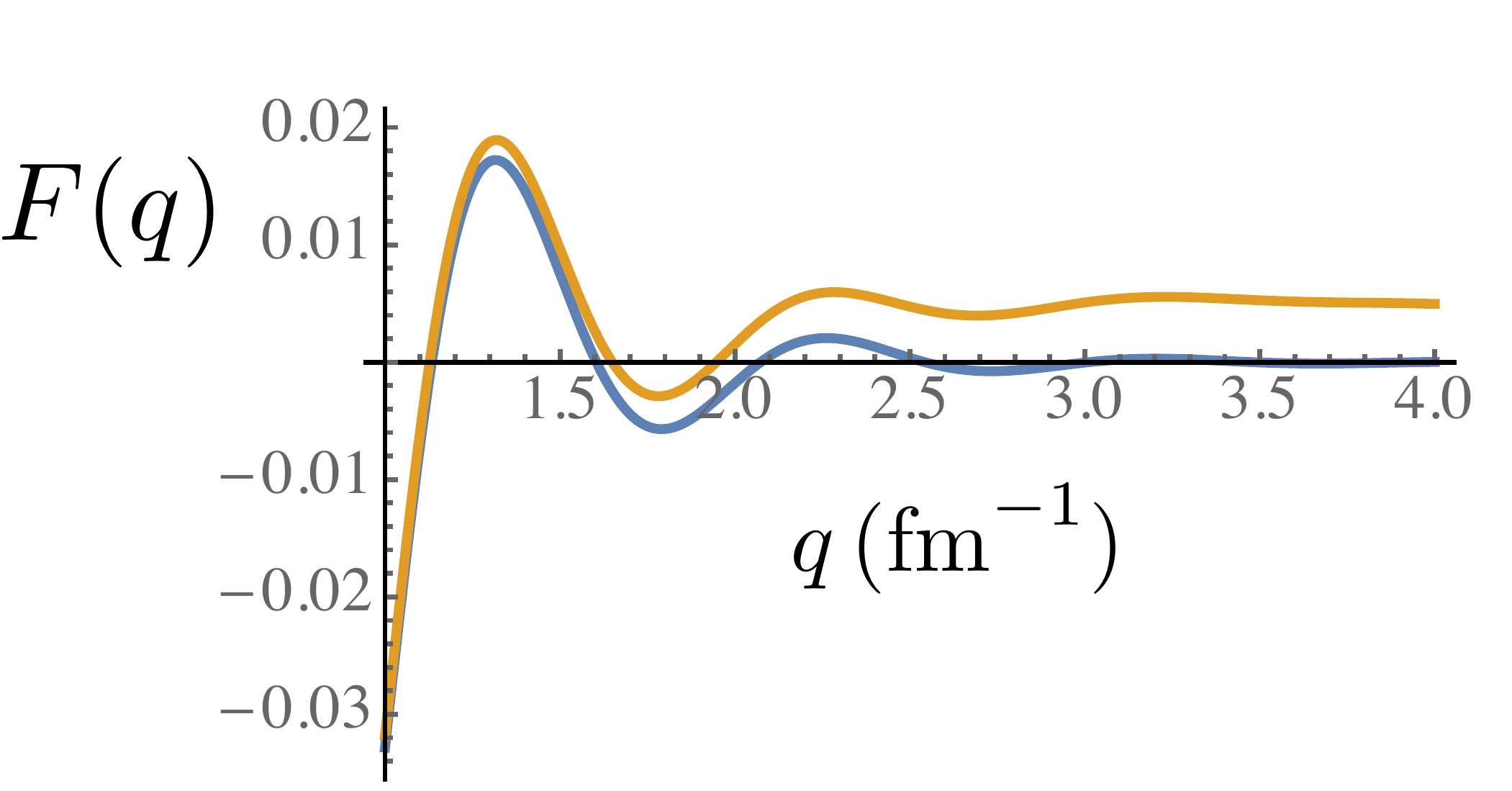}
 \caption{Nuclear Form Factor. Blue, solid, $F(q)$, Red, dashed,  $F(q)+\Delta F(q)$.  } \end{figure}

\section{ Two-Nucleon  Separation Distance   }


This section examines the physics of the two-nucleon separation distance. 
 Bound-state wave functions  are constructed using simple, two-parameter models of the $^3{\rm S}_1$ nucleon-nucleon interaction with parameters chosen to reproduce the measured scattering length and effective range  \cite{Brownjackson}.  As such, these are low-energy interactions.
These simple potentials contain features such as a hard core or Yukawa interaction that have been  parts of more realistic interactions.
 The range parameters of that reference are used here, with the strengths of the potential adjusted slightly so as to reproduce the value of the binding energy (2.2 MeV).
 The different potentials produce different bound-state wave functions and measurable differences are perceived through the behavior of the form factors (here the Fourier transforms of the square of the wave functions).
  The importance of the correction terms in the difference between using $\Oop_s$ and $\Oop$ is assessed. The scaling properties of the form factors are also presented in preparation for use in Sect.~V.
\\

\subsection{nucleon-nucleon hard core plus exponential potential}

 This potential is defined by having an infinite hard core at a separation $r_0$ and an attractive exponential
 potential $V(r)=-V_0\,e^{-(r-r_0)/a}$ ($V_0=1.92\,\fmi$) for larger separations. The model is exactly solvable. The values $r_0=0.4 $ fm and $a=0.45 $ \cite{Brownjackson} are used.   This potential (as  others in this section) is  a crude model for deuteron properties because there is no tensor force. 
\\

The $s$-state bound state wave function is determined by using the transformation $y=2a \g e^{-r/(2a)}$,  $\g=\sqrt{MB}$, where $B$ is the binding energy and $M$ the nucleon mass,  which converts the Schroedinger equation into Bessel's equation. Then  the  bound-state wave function is  
\bea u(r)= N\,J_{2a\g} (2a \sqrt{MV_0}e^{-r/2a} ),
\eea 
subject to the condition that $u(r_0)=0$. The factor $N$ is a normalization constant. One can check the large $r$ limit by using the small argument limit of the Bessel function ($J_\n(x)\sim x^\nu$) so that  $\lim_{r\to\infty} u(r)\propto e^{-\g r}$, as expected. 
The form factor of this model is the  bound-state matrix element of the operator 
\bea {\cal O}_\bfQ= e^{i \bfQ\cdot \bfr/2},\eea
in which the probe is defined to act only on one nucleon of the two-body system. Then the  form factor 
is given by
\bea &F(Q) ={2\over Q} \int_{r_0}^\infty  {dr\over r} {\sin{(Q/2\,r)} } u^2(r),\eea 
and can be re-expressed in terms of the momentum-space wave function $\psi(k)$
 given by 
 \bea \psi(k)={1\over \sqrt{2}\pi k}\int_{r_0}^\infty  dr \sin{k r}\; u(r),\eea
with \bea &F(Q) =\int d^3k\, \psi(k_+) \psi(k_-), \label{momgen}
\eea
and $\bfk_{\pm}\equiv \bfk\pm\bfQ/4 $.\\

If one uses the $V_{\rm low\,k}$ prescription of \eq{proj} one cuts off the 
momentum-space wave function at a relative momentum $\L$, with $\L=2.1$ fm$^{-1}$ a commonly used value.
The aim here is to see how much of the form factor (as a function of $Q^2$) is given by relative  momenta  that are greater than $\L$.  \\
 
 The cutoff form factor is then given by 
 \bea
F_\L(Q)= \int d^3k \psi(k_+) \psi(k_-)\Theta(\L-k_+)\Theta(\L-k_-).\nonumber\\
\label{fmom12} \eea
Using this form factor corresponds to using \eq{proj} for the wave function. Invariance of the form factor would be obtained if the probe operator were modified according to \eq{renorm} or \eq{fo}. The purpose in computing $F_\L(Q)$ is only to determine the values of $\L$ for which operator modification becomes necessary.\\
  
Fig.~2 shows  the form factor falling asymptotically as $1/Q^6$ and modulated by oscillations. Fig. ~3 shows the values of $\L$ necessary to achieve 5\%  accuracy in the form factor as a function of $Q$.   These are greater than 2.1 $\fmi$ for  values of $Q>2.2$ fm$^{-1}$, so such values of $Q$ require
 operator modification.  The use of \eq{fo} is not possible because of the hard core of the potential.

  \begin{figure}[h]\label{HCExpForm}
\includegraphics[width=4.9cm,height=4.249253cm]{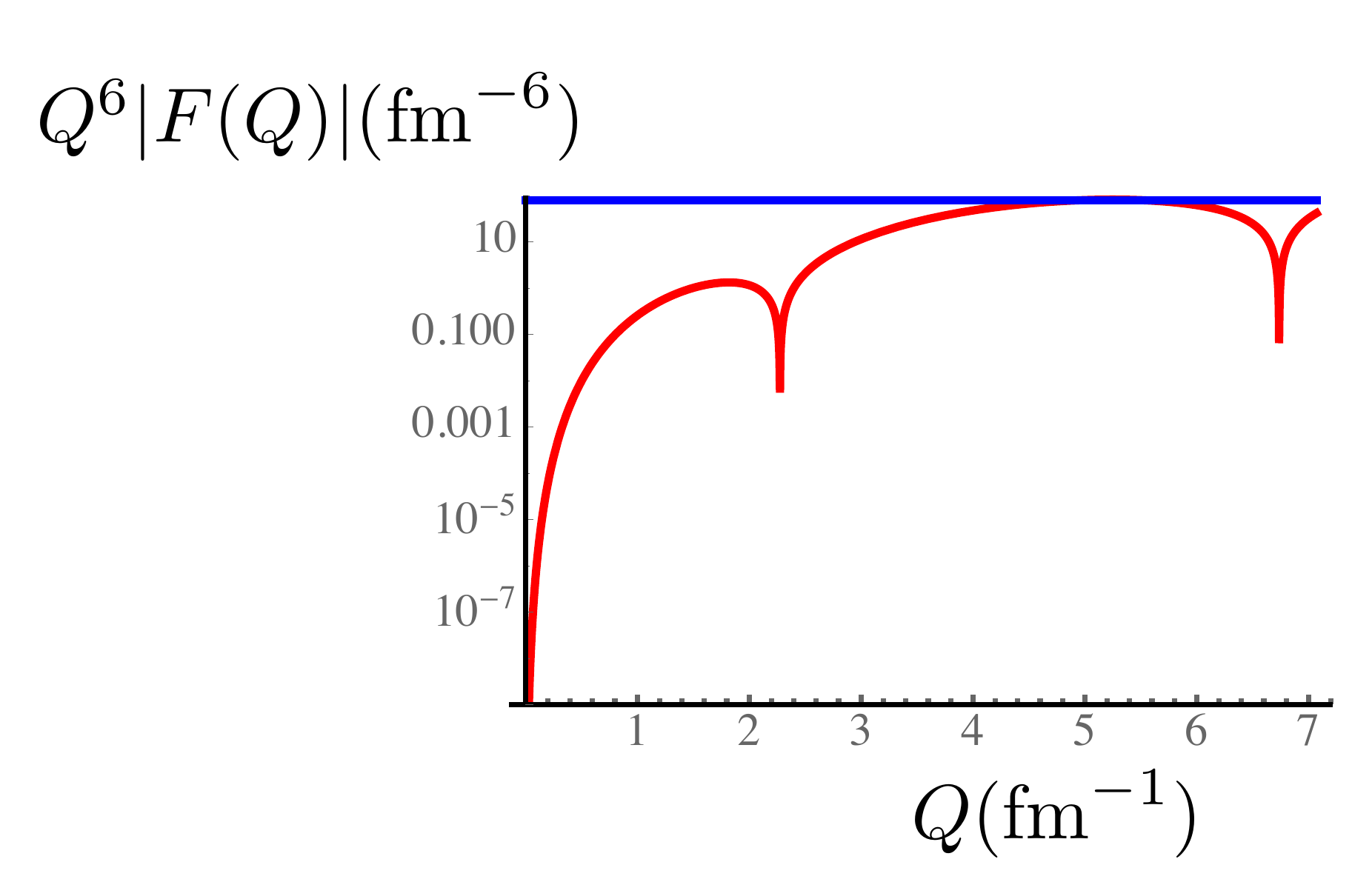}
 \caption{$F(Q)$  for hard core plus exponential potential } \end{figure}

 \begin{figure}[h]\label{hcexp}
\includegraphics[width=3.9cm,height=3.9253cm]{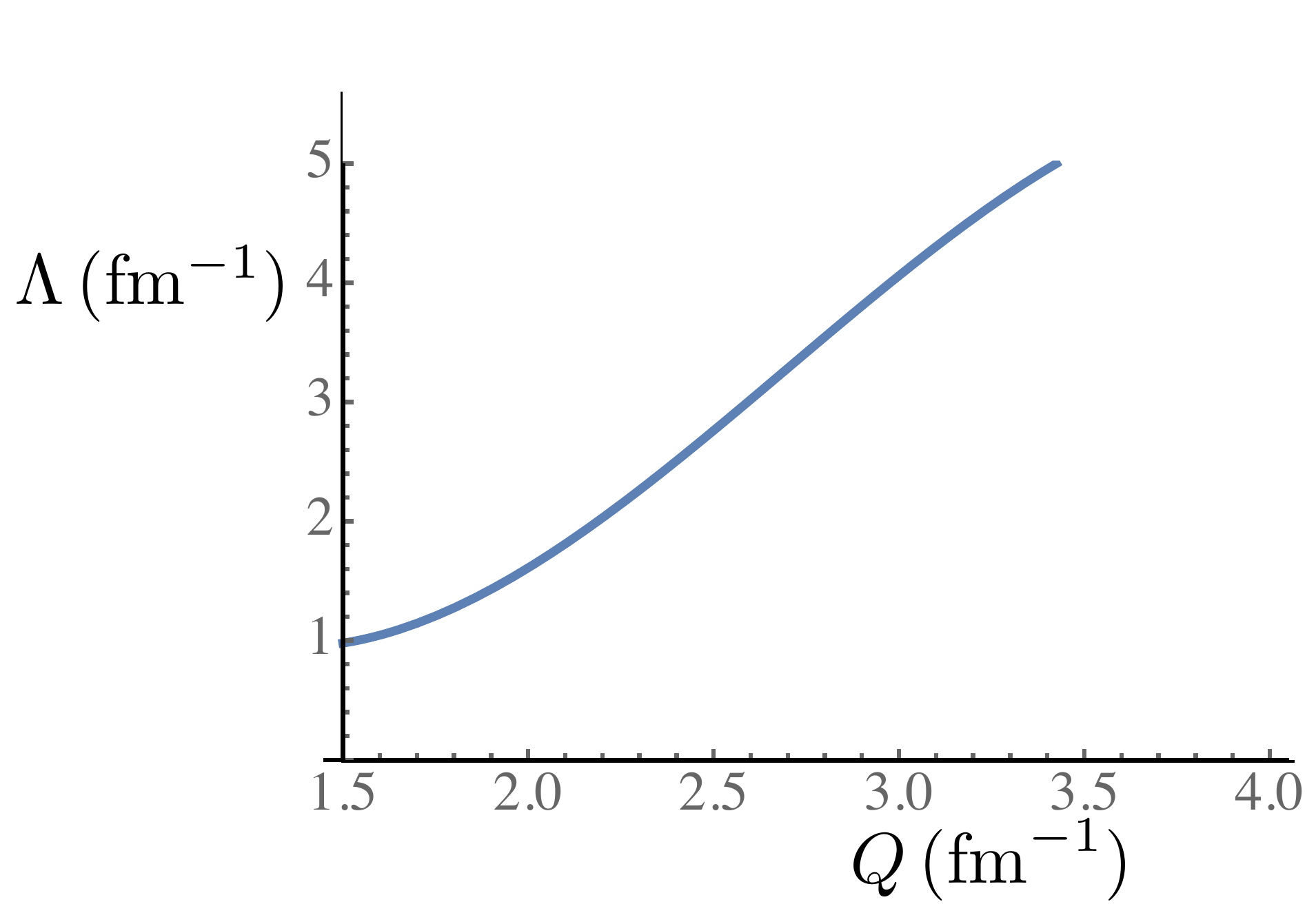}
 \caption{Value of $\L$ for which $F_\L(Q)/F_\infty(Q)=0.95$ as a function of $\L$.}\end{figure}



\subsection{Square well potential}

The next example is the square well potential with a radius of 2.205 fm  \cite{Brownjackson} and depth 0.157 fm$^{-1}$. 
The form factor  is shown in Fig.~4.    Fig.~5  shows the values of $\L$ necessary to achieve 5\%  accuracy in the form factor as a function of $Q$. 
 Operator modification is found to be important here for values of $Q>1.5$ fm$^{-1}$. The use of \eq{fo} is not possible because 
 the derivatives of the potential are delta functions.

  \begin{figure}[h]\label{sq}
\includegraphics[width=3.9cm,height=3.9253cm]{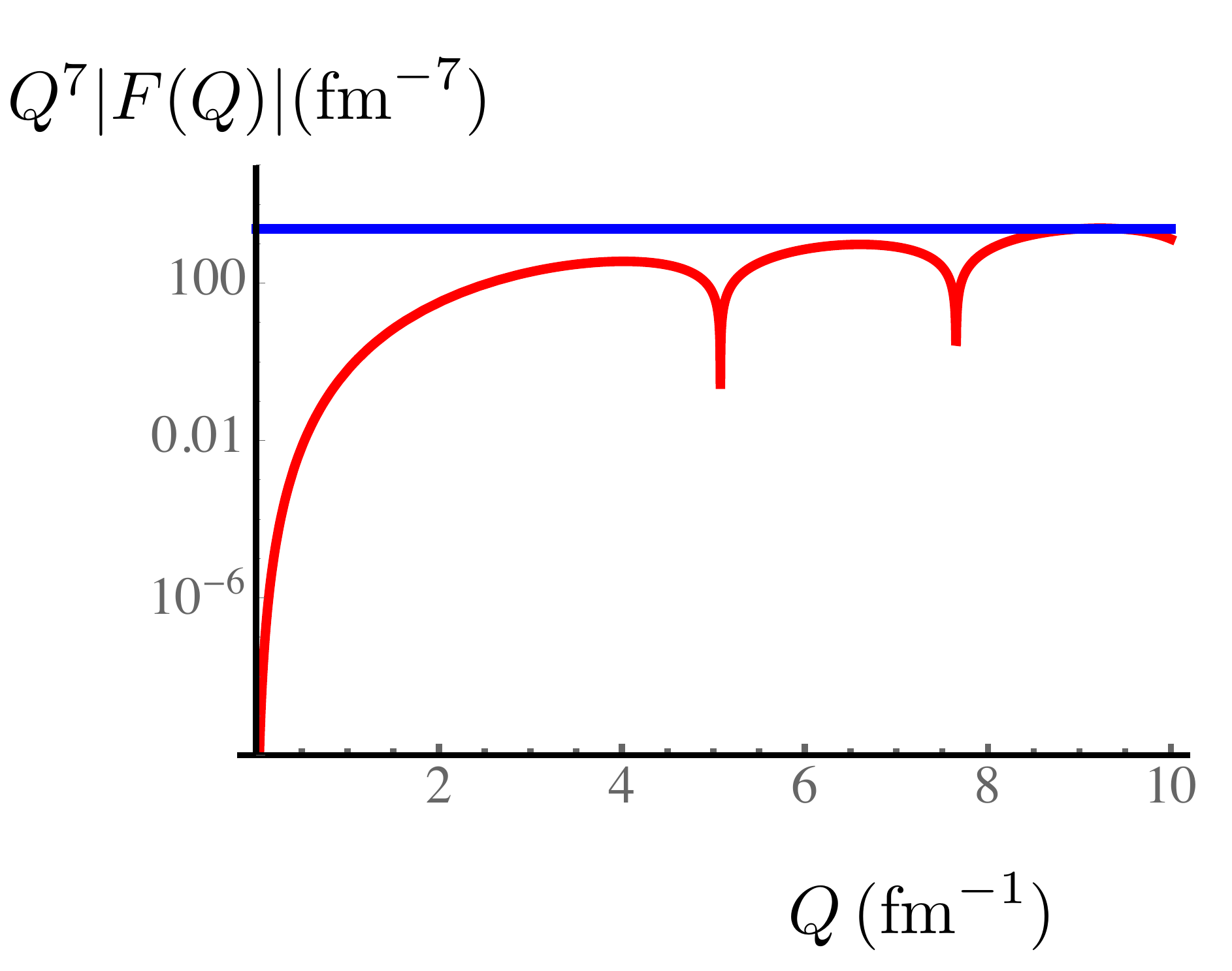}
 \caption{$F(Q)$  for square well potential } \end{figure}

 \begin{figure}[h]\label{sqcut}
\includegraphics[width=3.9cm,height=3.9253cm]{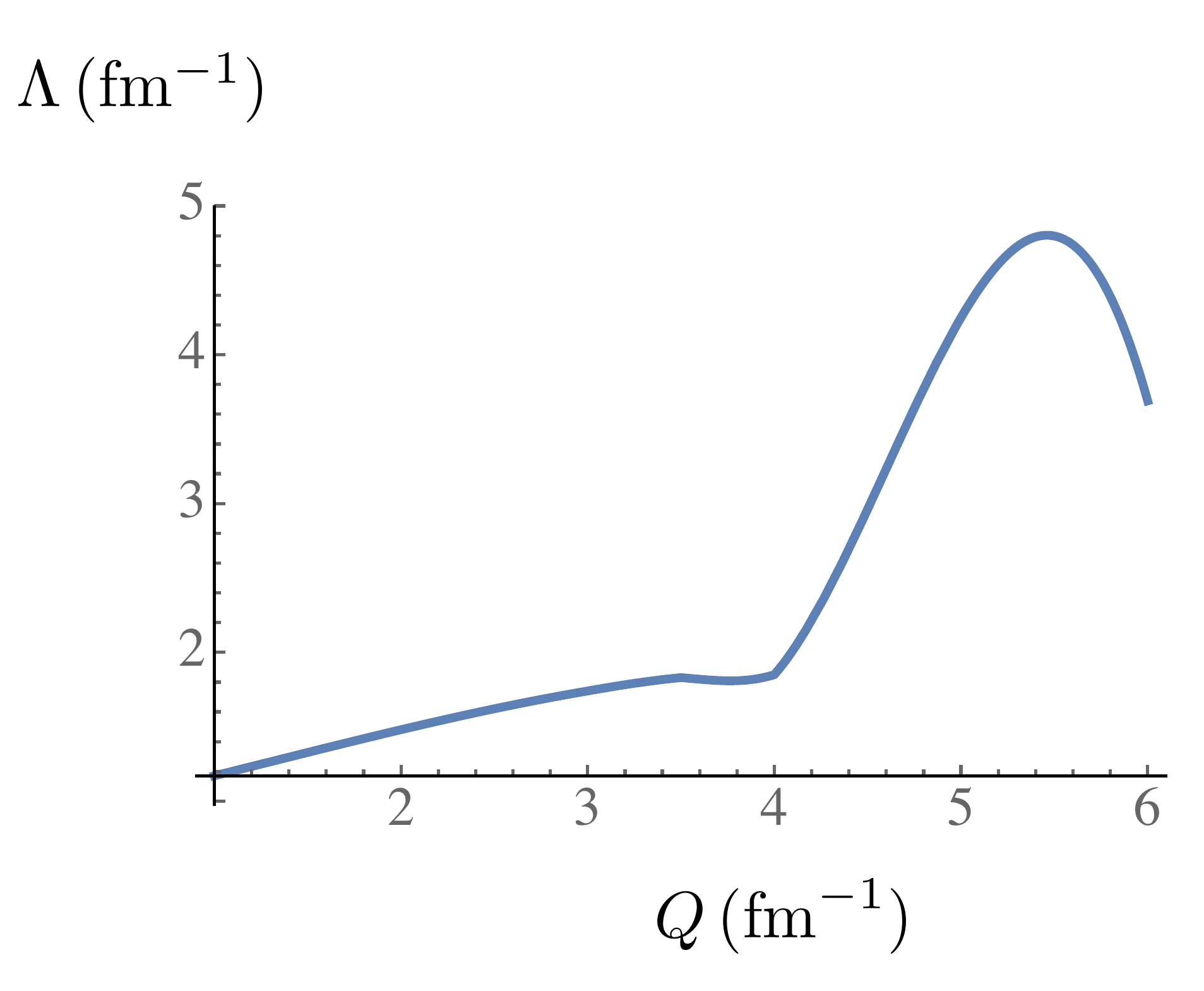}
 \caption{Value of $\L$ for which $F_\L(Q)/F_\infty(Q)=0.95$ as a function of $\L$ for square well potential. The rapid rise is due to a node in the form factor.} \end{figure}

\subsection{Exponential  potential}
The exponential potential is given by the expression $V(r)=-V_0 e^{-r/a}$ with  $a=0.76 $ fm \cite{Brownjackson} and $V_0=0.779 $ fm$^{-1}$.
 The form factor is shown in Fig.~6.   One sees that $F(Q)$ scales as $1/Q^6$.\\


 \begin{figure}[h]\label{E}
\includegraphics[width=5.9cm,height=4.53cm]{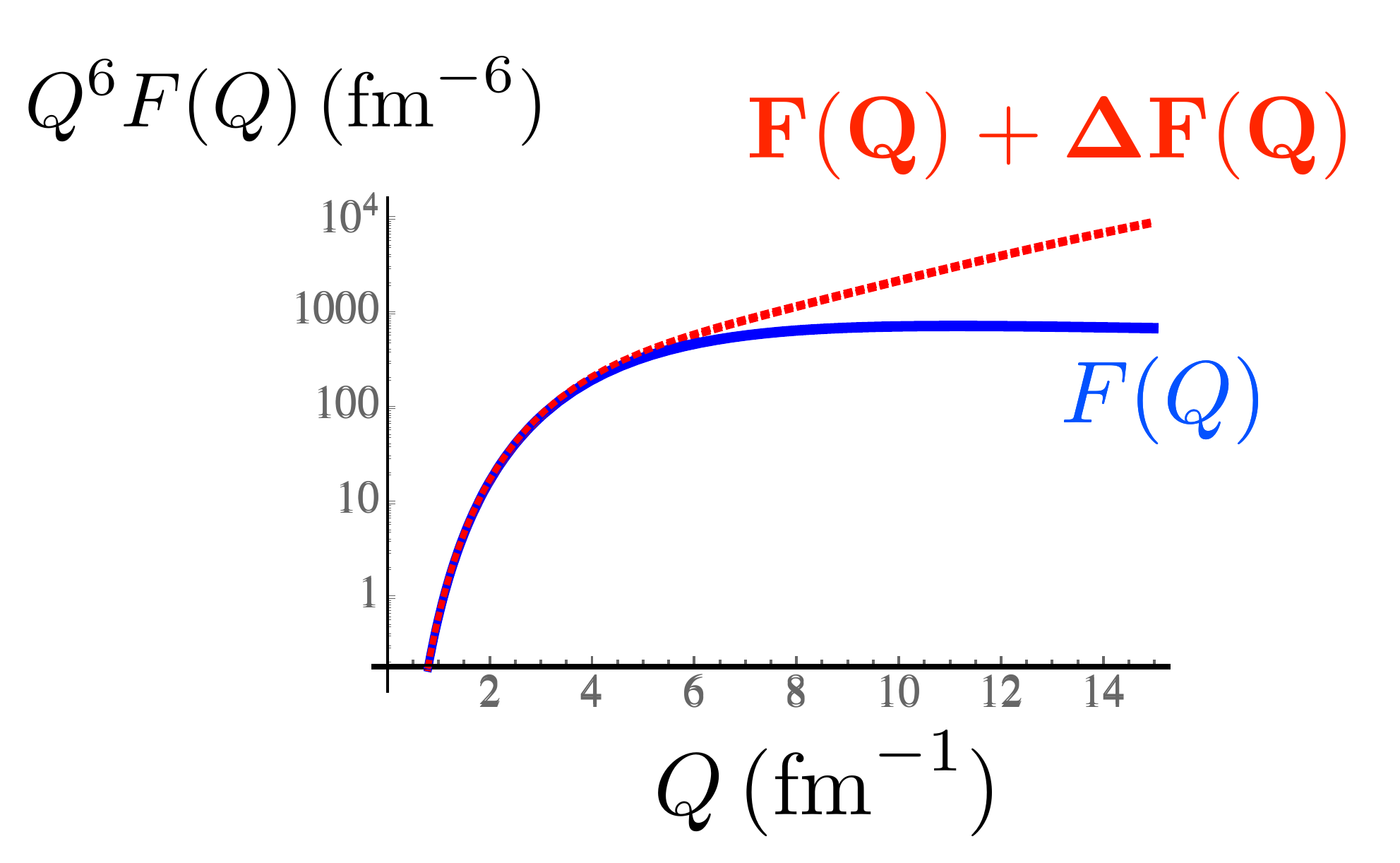}
 \caption{$Q^6F(Q)$  for exponential potential. Solid $F(Q)$.  Dashed $F(Q)+\D F(Q)$, see \eq{dfe}. } \end{figure}

This potential has well-defined derivatives so that one may use the probe operator evolution of \eq{fo} to study the change in the operator. For computing  the form
factor  of a two-body bound state  \eq{fo} becomes 
\bea\Oop_s\approx (1-{i\over 2} sM {dV\over dr}(\hat\bfr\cdot \bfQ))e^{i\bfQ\cdot\bfr/2}. \label{do}
\eea
The use of the second term of this equation causes a change to the computed form factor $\D F(Q)$ with
\bea& \D F(Q)={Q sM
\over 6}\int dr u^2(r) j_1(Qr/2) {dV\over dr}.\label{dfe}\eea
 The function $F(Q)+\D F(Q)$ is shown as the dashed curve of Fig.~6. One sees that  the term induced by the softening of the interaction causes a significant hardening of the  interaction starting for values of $Q$ as low as about $2\fmi$ and dominates for    $Q> 3 $ fm$^{-1}.$  If $\D F$ is large compared with $F$ it is necessary to include  higher-order terms in $s$, so the details would change.  Nevertheless, the Fig.~6 demonstrates the hardening of the probe interaction.


\subsection{Yukawa potential}
Here $V(r) =V_0 e^{-\m r}/r$ with $\m=0.411 \,{\rm fm^{-1}} $ as in   \cite{Brownjackson} and $V_0=0.25$.  The form factor, as shown Fig.~7,
scales as $Q^{-4}$. The   function $F(Q)+\D F(Q)$  (\eq{dfe}) is shown as the rising curve of Fig.~7. The dramatic  change in the probe operator is caused by the large derivative of the Yukawa potential at short distances.   The Fig.~7 again demonstrates the hardening of the probe interaction. \\
\begin{figure}[h]\label{Yuk}
\includegraphics[width=4.995cm,height=4.9953cm]{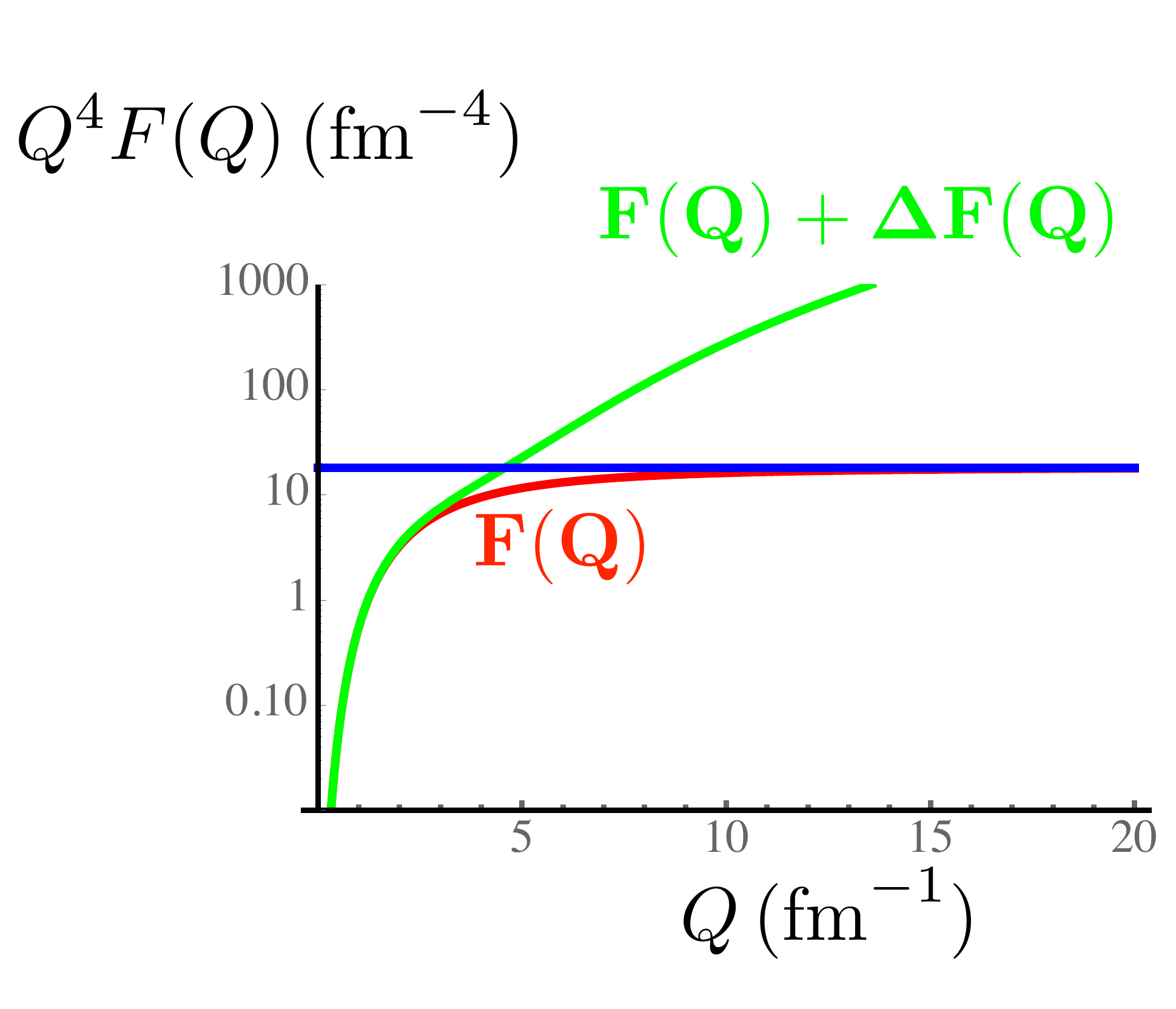}
 \caption{$Q^4F(Q)$  and $Q^4(F(Q)+\D F(Q))$ for  Yukawa    potential.}\end{figure}


\subsection{Influence of tensor force and higher $Q^2$}
The one-pion exchange potential (OPEP)  causes a tensor force that dominates 
the long  range properties of the deuteron. This  has been known since the discovery of the quadrupole moment of the deuteron in 1939. 
Furthermore, the OPEP by itself, along with a single parameter that provides a short-distance cutoff, is known to provide an approximate but  reasonable bound state wave function for the deuteron \cite{Friar:1984wi,Cooke:2001rq}.\\

The iteration of the tensor part of OPEP that occurs in solving the Schroedinger equation gives an S-state potential that acts approximately as an atractive delta function potential~\cite{,Kaiser:2001jx,Hen:2014lia,Hen:2016kwk}. This approximate delta function is the leading order term for the potential in  both EFT and pionless EFT.
In momentum space the S-state wave function has a node around  $k=2$ fm $^{-1}$, and the D-state dominates for $k$ between about 2 and 4 fm$^{-1}$ for many potentials that are in use in many-body calculations today.\\

The softening of the OPEP by the SRG means that the electromagnetic interaction must acquire  a  tensor force component. Including this effect in the probe operator  would add a
 complication.\\
 
 \subsection{Summary}

Softening of the NN interaction via a unitary transformation or projection operator procedure requires a corresponding transformation of interaction operators that increases their effects at high momentum transfer.   The examples shown indicate that  for some potentials the effects of transforming the operator are very important for momentum transfers greater than about 5 fm$^{-1}$, an important region for current experiments that attempt to discover new phenomena. Furthermore, the transformed operators  cannot be obtained easily for  some potentials.\\

The use of the impulse approximation that involves using bare, untransformed operators simplifies the interpretation of experiments and therefore seems best suited for discovery purposes.

\section{Two-nucleon separation in nuclei:  observing high momentum and  short-distance features. }

The previous Section discusses how high momentum components may arise from  interactions between nucleons. The present Section is concerned with the manifestation of such effects in nuclei,  and also one way to observe the relation between short-distance and high momentum physics.\\


Bethe \cite{PhysRev.103.1353}   wrote that ``Indeed, it is well established that the forces between two nucleons are of short range, and of very great strength"  and
``there are strong arguments to show that the
two-body forces continue to exist inside a complex
nucleus ".\\

 Brueckner, Eden, and Francis, \cite{PhysRev.98.1445} used a variety of nuclear reactions to argue that the nuclear wave function contains nucleons 
 with a significant probability to have high momentum.  One particularly telling example is the significant cross sections observed in the $(p,d)$ reaction with 95 MeV protons. The neutron in the nucleus must have high momentum comparable to that of the proton,  about 420 MeV/c,
so that combination with the incident proton allows the deuteron to emerge from the nucleus. The only way a bound neutron could acquire such momentum is via interactions with another nearby nucleon.\\

Bethe continued ``All these processes show that the `potential' is fluctuating violently
from point to point in the nucleus, which is compatible with the assumption that two-body forces continue to act inside the nucleus without much modificcation."
The idea of two strongly interacting nucleons, acting independently of the other nucleons (the independent pair approximation)  is the basis of Bruckner theory ~\cite{PhysRev.97.1353} which provided a fundamental explanation of how nuclear saturation and the shell model of nuclei arise from fundamental, hard, short-ranged interactions of nucleons.  This means that the nucleon-nucleon separation distance is related, via the nucleon-nucleon interaction,  to the size of the entire nucleus.\\

One  modern implementation of the independent pair approximation is the generalized contact formalism (GCF)
\cite{Cruz-Torres:2019fum}.
The GCF  is an effective model that provides a factorized approximation for the short-distance (small-$r$) and high-momentum (large-$k$) components of the nuclear many-body wave function. Its derivation relies on   the strong relative interaction of closely separated  nucleons 
 and their weaker interaction with the residual $A-2$ nuclear system~\cite{Weiss:2015mba,Weiss:2016obx,Cohen:2018gzh}. Using this approximation, the two-nucleon density in either coordinate or momentum space (i.e., the probability of finding two nucleons with separation $r$ or relative momentum $q$) has been  expressed at small separation or high momentum as~\cite{Weiss:2016obx}:
\begin{align}
	\rho_A^{NN,\alpha}(r) & = C_A^{NN, \alpha} \times |\varphi_{NN}^{\alpha}(r)|^2 , \nonumber \\
	n_A^{NN,\alpha}(k) & = C_A^{NN, \alpha} \times |\varphi_{NN}^{\alpha}(k)|^2 , 
\label{Eq1}
\end{align}
where $A$ denotes the nucleus, $NN$ denotes the nucleon pair being considered ($pn$, $pp$, $nn$), and $\alpha$ stands for the nucleon-pair quantum state (spin 0 or 1). $C_A^{NN, \alpha}$ are nucleus-dependent scaling coefficients, referred to as ``nuclear contact terms'', and $\varphi_{NN}^{\alpha}$ are   two-body wave functions that are given by the zero-energy solution of the two-body Schr\"odinger equation for the $NN$ pair in the state $\alpha$. The functions $\varphi_{NN}^{\alpha}$ do not depend on the nucleus, but do  depend on the $NN$ interaction. \\

The authors \cite{Cruz-Torres:2019fum} state that 
an important feature of the GCF is the equivalence between short distance and high momentum, which  is built into \eq{Eq1}  by using  the same contact terms $C_A^{NN, \alpha}$ for both densities.   This equivalence is established by   extracting the contacts separately from the coordinate- and momentum-space nuclear wave functions.  The present section is devoted to finding a  direct correspondence between short distance and high momentum.\\


This analysis uses the  zero-energy  Lippmann-Schwinger (LS)  equation  and asymptotic expansions obtained by integration by parts~\cite{Erdelyi}.
The  LS equation for scattering at 0 energy is given by 
\bea 
\varphi_{NN}^{\alpha}(k)= {-M\over k^2} \int {d^3r\over (2\pi)^{3/2}} e^{-i\bfk\cdot\bfr}  V(r) \varphi_{NN}^{\alpha}(r).\label{LS}\eea
If the potential is an approximate  delta function in coordinate space, then 
$
\varphi_{NN}^{\alpha}(k)\sim {1\over k^2} .
$\\

For other interactions it is useful to express the  $S$-wave, momentum-space wave function as:
\bea 
\psi(k)=- {M\over \sqrt{2}\pi k^3}\int_0^\infty dr \sin({k r)} \ V(r)\  u(r),\label{ls1}\eea
where $u(r) $ is the S-state radial wave function and in which the labels $NN,\a$ are suppressed.
One derives expansions for asymptotic values of the momenta by replacing the $\sin(kr) $ appearing in the integral of \eq{ls1} by ${-1\over k}{d\cos kr\over dr} $. Then one can get higher-order terms by writing  $\cos(kr) ={1\over k} {d\sin{kr}\over kr}$. The result, defining $K\equiv {M\over \sqrt{2}\pi}$, assuming that  the potential is not a delta function, and that $Vu$ and its derivatives exist at $r=0$  is:
\bea &
\psi(k)={K\over k^4}\int_0^\infty dr {d\cos kr\over dr} V(r) u(r)\\&={K\over k^2}[-V(0) u(0) -\int_0^\infty dr \cos kr (Vu)']\nonumber
\\&
={K\over k^4}V(0) u(0)+{K\over k^6}(Vu)''(0)+{K\over k^8}(Vu)''''(0)+\cdots\label{ae}
\eea
If the potential is non-local of the form $V(r,r')$ the product $Vu$ in \eq{ls1} is replaced by  \bea V_u(r)\equiv r\int_0^\infty dr' V(r,r') u(r')\eea  and  the derivatives thereof that appear in \eq{ae} are replaced by derivatives of $V_u$ at the origin.\\

One may  classify the asymptotic behavior  obtained from different classes of potentials.
\begin{itemize}
\item Class I: The potential is a delta function. Then  $\psi(k)\sim {1\over k^2}$ as in leading-order pion-less effective field theory. Or as in Ref~\cite{RevModPhys.92.025004}  showing that an approximate delta-function potential arises from treating the  iterated effects of the one pion exchange potential.

\item Class II:  $u(0)=0$ but $V(0) u(0)\ne0. $ An example is $V\sim 1/r $  and $u(r) \sim r$ for small values of $r$. In this case, $\psi(k)\sim {V(0)u(0)\over k^4}$

\item Class III: $u(0)=0$, $V(0) u(0)=0. $ An example is  the exponential potential for which $V(0)\ne0 $ is finite and $u(0)=0$.  In this case,  $\psi(k)\sim {V'(0)u'(0)\over k^6}$

 \item Class IV: The potential has a  hard core  potential, infinitely repulsive for a distance less than a core radius, $r=c$. Then using  $u(c)=0$, $u'(c)\ne0$ and taking the Fourier transform of the wave function: 
 \bea &
\psi(k)={-K\over k^2}\int_c^\infty dr {d\cos kr\over dr} u(r)\nonumber\\&  
\sim{K\over k^3}\sin(k c) u'(c).\label{hcee}\eea

\item  Class V:  $Vu$ and all of its derivatives vanish at the origin. This is the square well of range $R$. Then using  the LS equation  yields 
\bea &\psi(k) 
\sim {KV(0)\over k^4} \cos(k R)u(R).\eea
 
 \item Class VI: Non-local potentials. The quantity $\psi(k) \propto \lim_{r\to0}(V_u(r))/k^4$  unless the limit vanishes.  The 
 Yamaguchi potential ~\cite{PhysRev.98.69} $V(r,r') \propto {e^{-\m r}\over r}{e^{-\m r'}\over r'}$ provides an example of $\psi(k)\propto {1\over k^4}$. A power  law fall-off would be obtained even if previous limit did vanish because some non-zero  even-numbered derivatives of $V_u$ at the origin must occur. 

 \end{itemize}

In each of the first   five cases the product of the potential and wave function at short separation distances  determines the  high-momentum behavior of the momentum-space wave function. For non-local potentials  the high-momentum behavior is controlled by $V_u$ and/or its derivatives at the origin. Once, again short-distance behavior determines the high momentum content. Moreover, in each case there is a power law fall-off with increasing $k$. This slow fall with increasing $k$ means that significant high-momentum content can be expected for all of  the interactions of 
Classes I through VI. \\

A power-law fall off can be uniquely avoided if the potential is a function of $r^2$. In that case, all of the terms in the series of \eq{ae} would vanish because of the vanishing of all odd-number derivatives of $V(r^2)$ at the origin. No realistic nucleon-nucleon  potential in current use is a function of $r^2$. This means that significant high momentum content can be expected.

\subsection{Form factors at high momentum transfer}

The previous analysis of zero-energy wave functions is also applicable to bound-state wave functions. For a binding energy $B$ the $-M\over {k^{2n}}$ factors of \eq{ae} is replaced by $-{M\over (k^2 +MB)k^{2n-2}} \approx  -{M\over k^{2n}}$ in asymptotic expansions.\\

An approximate relation between the momentum space wave function and the elastic form factor can be obtained using \eq{momgen}. Ref.~\cite{Brodsky:1989pv}
argued that the dominant  contributions to the integral occur when $\bfk=\pm \bfQ/4$. Then
\bea F(Q)\propto\psi(Q/4).\label{wow}\eea This result depends on factorizing the momentum dependence of the potential, $\tilde V$  from that of the wave function, and is denoted the factorization approximation. The procedure is to 
 use the LS equation to represent the wave functions appearing in \eq{fmom12}. Then \eq{wow} emerges if  \bea \int d^3k \tilde  V(|\bfQ/4-\bfk|)\psi(k)\approx \tilde V(Q/4)\int d^3k\psi(k).\label{fac}\eea
 The integral over $d^3k$ is the wave function at the origin of coordinate space.
 \\

The result \eq{wow} is remarkable. It means that under certain conditions, in principle, it is possible to measure the wave function of a system, or at least its momentum dependence in a specific regime. This means that general statements about the unmeasurable nature  of  wave functions are not correct.\\

An  (unrealistic) experiment  in which one could attempt to test \eq{wow} is elastic electron scattering from a $b\bar b$ meson. Elastic scattering on the deuteron  is complicated by the need to include the effects of meson exchange currents and corrections to the non-relativistic treatment~\cite{Marcucci:2015rca}.   Calculations of deuteron form factors for momentum transfers greater than about 7 fm$^{-1}$ are not shown in that review. \\

Note also  that nucleon-nucleon scattering at laboratory energies less than 350 MeV  does not yield significant constraints on $ \tilde V(Q/4)$  for large values of $Q$~\cite{PhysRevC.69.044004}. Large momentum transfer means that large kinetic energy is needed.\\

The following text explains how the different classes of potentials discussed here  can be  or cannot be manifest by measurements of form factors  as expressed in  \eq{wow}.\\

Class I: $V$ is a delta function in coordinate space, and therefore a constant in momentum space. The wave function $\psi(k)$ is mainly determined by the propagator in which $k$ and $Q$ of \eq{fac} of the same importance. The factorization argument does not apply. \\

Class II: The Yukawa potential $V(r)=V_0 e^{-\m r}/r$. The product $Vu$ is well defined as $r\to0$, because then $u(r)\propto r$. Thus \eq{ae}  predicts $\psi(Q)\sim 1/Q^4$ and the form factor show in  in Fig.~7 also shows a $1/Q^4$ behavior. \\

Class III: The exponential potential.  In accord with \eq{ae}  the wave function falls as $1/Q^6$, and so does the form factor shown in Fig.~6.\\

Class IV: Hard core plus exponential. Fig.~2  shows  oscillations expected from \eq{hcee}  but  factorization does not work because the discontinuity of $u'(r)$ at $r=r_0$ induces  large momentum components. \\

Class V: Square well potential. The factorization approximation is not accurate, although oscillations with period $(2\pi/R\approx 3 \,{\rm fm}^{-1}$ are seen. This is because condition of \eq{fac} are not maintained due to oscillations that cause 0's in $\tilde V$ for large values of the argument.\\

In summary,  the short distance behavior of the potential times the coordinate-space radial  wave function determines the high momentum dynamics in all cases. If the factorization approximation of \eq{fac} is valid and the probe operator is well-known  the measurement of the form factor determines the high-momentum behavior of the wave function.
 

\section{The $(e,e'p)$ reaction:  discovery at the nucleon-nucleon separation scale}

The $(e,e'p)$ reaction  occurs if an  electron knocks out a nucleon so that an initial nuclear state of $A$ nucleons is converted to a final nuclear state of $A - 1$ nucleons. \\
 
In the plane wave impulse approximation (PWIA), an electron transfers a single virtual photon with momentum $\bfq$ and energy $\nu$ to a single proton, which then leaves the nucleus without interacting with another nucleon on the way out of the nucleus, see Fig. 10.
There are various corrections- final state interactions, meson exchange currents etc. However, one can account for such effects by using appropriate kinematics and including the effects of final state interactions, see {\it e.g.}~\cite{Schmidt:2020kcl}. \\

For high-momentum transfer processes the outgoing nucleon has high energy, greater than the 350 MeV that is used to constrain nucleon-nucleon potentials.
The softening effects of unitary transformations on nucleon-nucleon potentials  requires that the potential be Hermitian. No realistic Hermitian potential applicable for scattering energies greater than about 1.5 GeV exists at the present time.  
 This means applying a  unitary transformation to soften the interaction is not practical.  Instead, the final state interactions can be treated using the Glauber approximation in which the nucleon-nucleon scattering cross sections are used as input to form the optical potential \cite{Glauber:1970jm}.\\

\begin{figure}[h]\label{PWIA}
\includegraphics[width=3.5cm,height=3.53cm]{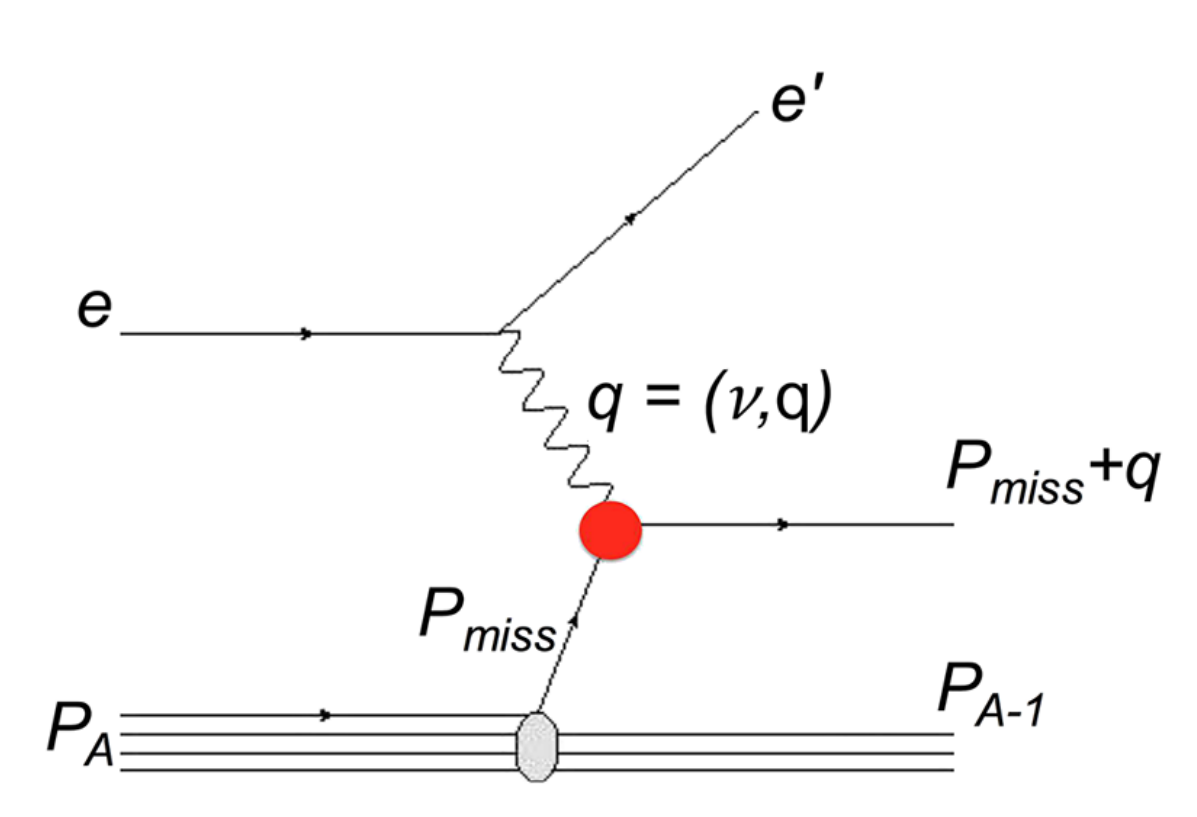}
 \caption{ A nucleus emits a nucleon of four-momentum $P_{\rm miss}$ that absorbs a virtual photon of four-momentum $q$ to make a final-state nucleon of four-momentum $P_{\rm miss}+q$, with $(P_{\rm miss}+q)^2=M^2$, where  $M$ is the nucleon mass.}\end{figure}

If the background effects mentioned above are handled correctly,  the scattering amplitude is proportional to the wave function of the struck bound nucleon~\cite{Walecka:1995mi}:
\bea {\cal M}\propto \psi(P_{\rm miss}).\label{imp}\eea

Once again (as in \eq{wow}) the scattering amplitude is seen to directly accesses information about the momentum dependence of the wave function. 
This feature has enabled experimental  studies to show that the high momentum part of the wave function is dominated by short-range correlations 
(SRCs)~\cite{Fomin:2017ydn}. These are 
pairs of nucleons with large relative and individual momenta and smaller center-of-mass (c.m.) momenta, where large is measured relative to the typical nuclear Fermi momentum $k_F \approx 250\,\rm MeV/c$~\cite{Hen:2016kwk,Atti:2015eda}. At momenta just above $k_F$ $(300 \le k \le 600\,\rm MeV/c)$, SRCs are dominated by $pn$ pairs~\cite{Tang:2002ww,Piasetzky:2006ai,Subedi:2008zz,Korover:2014dma,Hen:2014nza,Duer:2018sby,Duer:2018sxh}. This $pn$ dominance is due to the tensor part of the nucleon-nucleon ($NN$) interaction~\cite{schiavilla:2006xx, Alvioli:2007zz}.\\

The presence of  nucleon-nucleon short ranged correlations in nuclei  has many  implications for the internal structure of nucleons bound in nuclei~\cite{Hen:2013oha,Hen:2016kwk,Schmookler:2019nvf}, neutrinoless double beta decay matrix elements~\cite{Kortelainen:2007rh,Kortelainen:2007mn,Menendez:2008jp,Simkovic:2009pp,Benhar:2014cka,Cruz-Torres:2017sjy,Wang:2019hjy}, nuclear charge radii~\cite{Miller:2018mfb}, and the nuclear symmetry energy and neutron star properties~\cite{Li:2018lpy}.\\

If SRG transformations are applied to the strong-interaction Hamiltonian, the necessary use of hardened interactions (discussed in Sect.~V) in analyzing experiments would complicate their interpretation.\\



\section{Virtuality -a small-distance scale  }

Bound nucleons (of four momentum $p$) do not obey the   standard Einstein relation $p_\m p^\m=M^2$, and are said to be off the mass shell. The average binding energy is much, much less than the nucleon mass, so the violation of the Einstein relation can be ignored when computing or understanding  many average nuclear properties. \\

If one looks in more detail and examines  nucleon-nucleon scattering, one sees that  the intermediate nucleons must be  off their mass shell. In the Blankenbecler-Sugar \cite{PhysRev.142.1051} and Thompson reductions~\cite{Thompson:1970wt} of the Bethe-Salpeter equation~\cite{Salpeter:1951sz} one nucleon emits a meson of 0 energy and non-zero momentum and the other nucleon absorbs the meson. Since the momenta of the nucleons have changed, but their energy hasn't changed,  the intermediate nucleons are off their mass shell. In other reductions of the Bethe-Salpeter  equation \cite{Gross:1969rv}, one nucleon is on the mass shell, and the other is not.  This means that the nuclear wave function, treated relativistically, contains nucleons that are off their mass shell. Such nucleons must undergo interactions before they can be observed, and are denoted as virtual. The difference $p^2-M^2$ is related to  the virtuality~\cite{Miller:2019mae}.\\

Experiments~\cite{Egiyan:2003vg,Egiyan:2005hs,Fomin:2011ng} using leptonic probes at large values of Bjorken $x$ interrogate the virtuality of the bound nucleons. To see this, 
consider the PWIA situation with $(P_{\rm miss}+q)^2=M^2$, let $q$ have the four-momentum $(\n,{\bf0}_\perp,-\sqrt{\n^2+Q^2}))\approx (\n,{\bf0}_\perp,-(\n+{Q^2\over 2\n})$, in the Bjorken limit with $q^2=-Q^2$, $Q^2\to\infty,\,\n\to\infty$, and  $Q^2/\n$ finite. Then with $q^-=q^0-q^3\approx 2\n\gg q^+\approx - M x$, $x={Q^2\over 2M \n}$,
one finds that

\bea {\cal V} \equiv  {\Pm^2-M^2\over M^2} \approx -{Q^2\over M^2}({\Pm^+\over M}-1 ).\label{v0}\eea
This quantity ${\cal V}$, defined here as  the virtuality, is generally not zero. For example, experiments have been done with $Q^2=3 $ GeV$^2$, ${\Pm^+\over M}=1.5$ for which  $ {\cal V} \approx -1.5$. Plateaus, kinematically  corresponding to to scattering by  a pair of nucleons, have been observed \cite{Fomin:2017ydn} in this region.
Treating highly virtual nucleons requires including relativistic effects. A recent study is~\cite{Weiss:2020mns}.    
\\

 The only way for a nucleon to be so far off the mass shell is for it to be interacting strongly with another nearby nucleon. To see that,  consider a configuration of two bound nucleons, initially at rest in the  nucleus. This is a good approximation for roughly 80\% of the nuclear wave function. To acquire   the large missing momentum of the previous paragraph, one nucleon must exchange a boson or bosons with four-momentum comparable to that of the incident virtual photon as shown in Fig.~10.
\begin{figure}[h]
\includegraphics[width=2.9cm,height=2.5053cm]{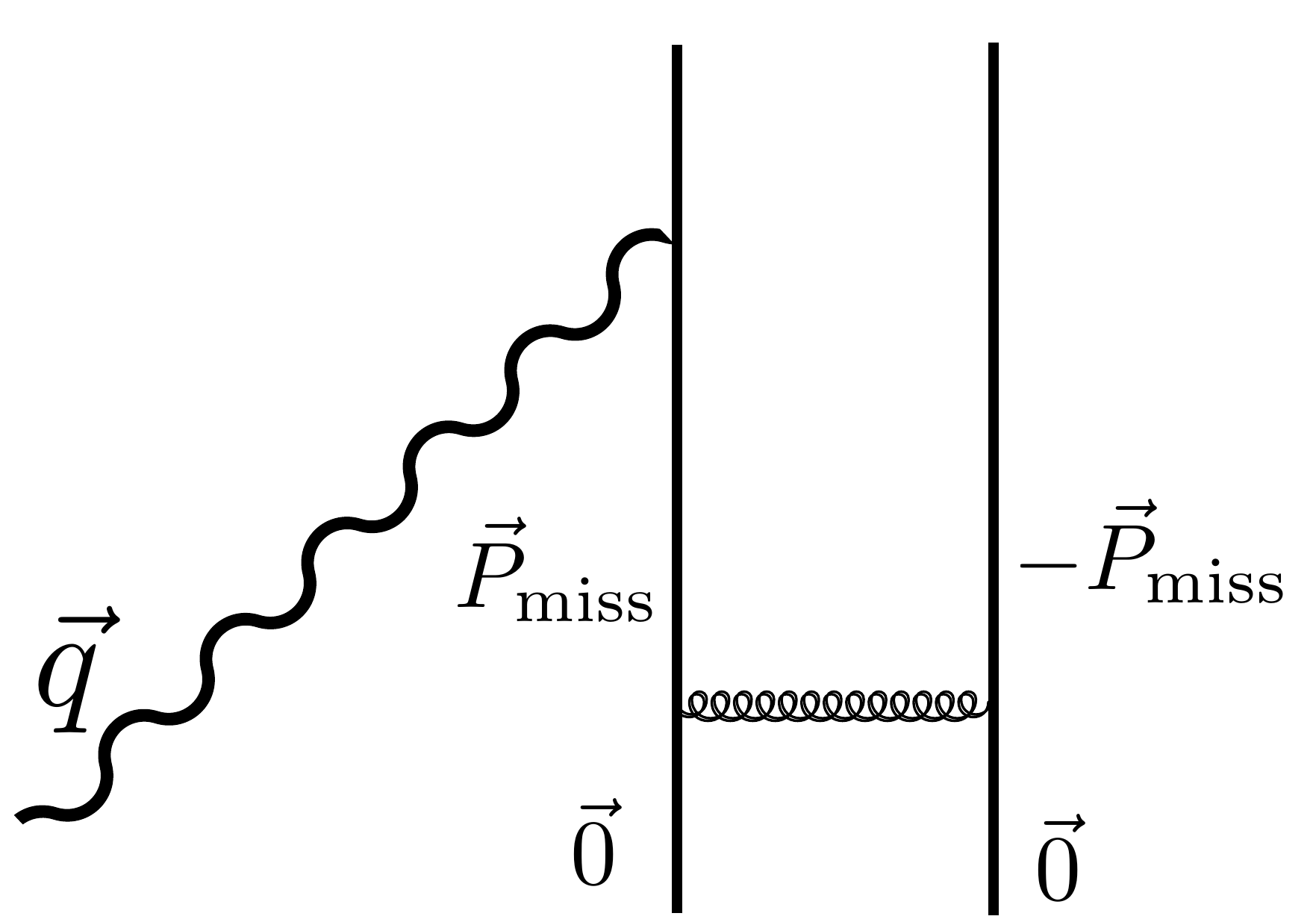}
 \caption{The  strong interaction represented by the wiggly line exchanges a momentum $\vec P_{\rm miss}$ between two nucleons. }\end{figure}
Such a bosonic system can only travel a short distance $\D r$ between the nucleons with 
\bea \D r\sim{1\over |\vec P_{\rm miss}|}   .\label{small}\eea
 Thus a highly  virtual nucleon gets its virtuality from another nearby nucleon which must be  closely separated.  High virtuality is a short-distance phenomenon.
  As such, it serves as an intermediate step between using nucleonic  and quark degrees of freedom \\

Ref.~\cite{PhysRevLett.123.042501} attempted to find a difference between the effects of  highly virtual nucleons and the effects of high local density.  The simple arguments presented here show that there is a direct connection between high local density and high virtuality. It is therefore not possible to distinguish the two effects. This issue is discussed in more detail in Ref.~\cite{Hen:2019jzn}.\\

 In evaluating Feynman diagrams the  lowest-order effects of the non-vanishing of  $\cal V$ can be cancelled by propagators and re-organized into low energy constants  See Fig.~10. But understanding the fundamental origin of virtuality  would allow a deeper understanding of nuclear physics. \\

\begin{figure}[h]\label{CD}
\includegraphics[width=5.9cm,height=2.5053cm]{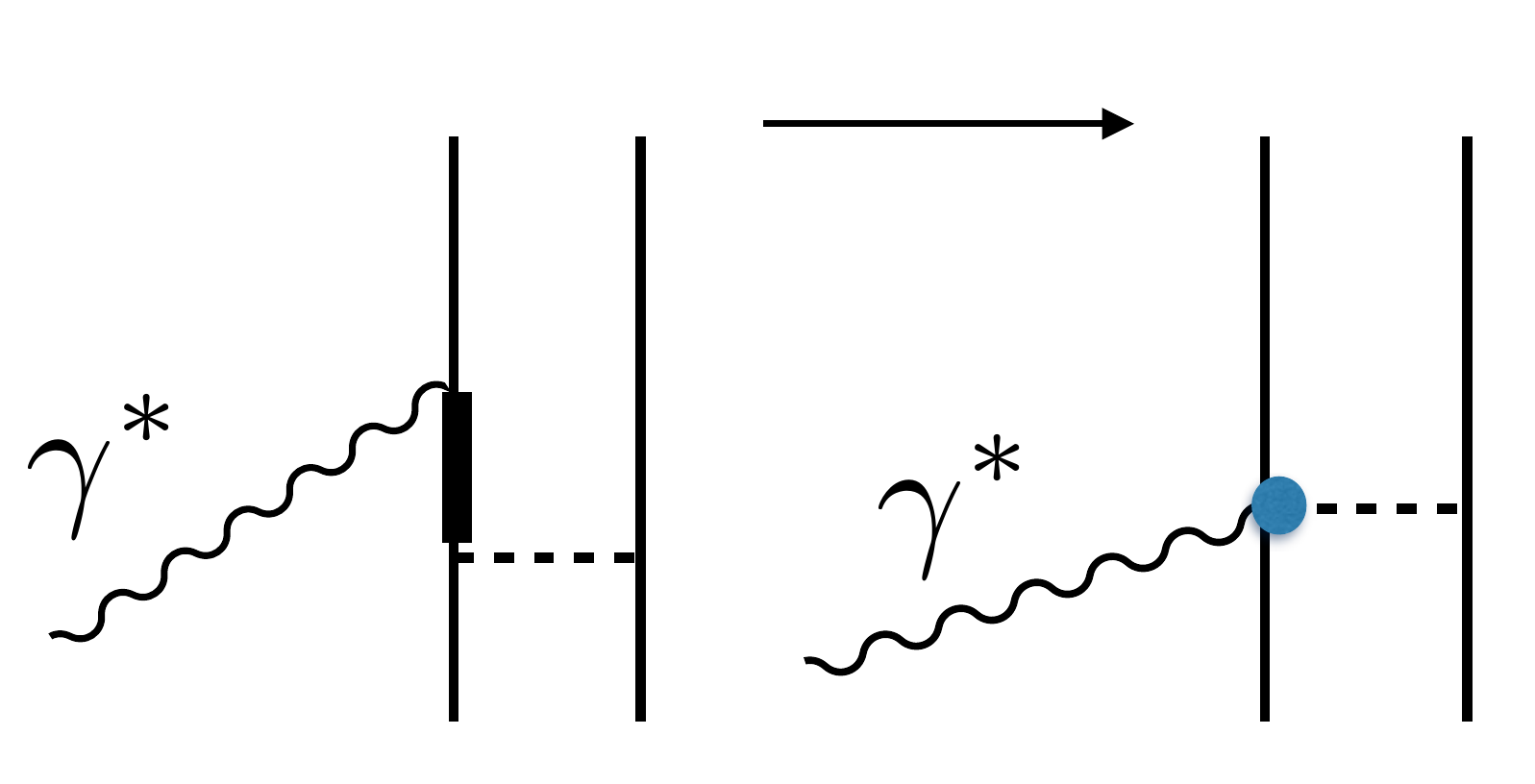}
 \caption{The effects of a virtual intermediate nucleon (indicated by the heavy line)  may be replaced by using a two-nucleon contact interaction.}\end{figure}

To better understand the connection between virtuality and quark degrees of freedom, consider a virtual nucleon as a superposition of physical states that are eigenfunction of the QCD Hamiltonian.
Virtual states with nucleon quantum numbers can be expressed using the completeness of states of QCD:
\bea |N(\Cv)\ra=\sum_{n=1}^{n_{\rm max}} c_n |N_n\ra, \label{sup} \eea in which 
the states $|N_n\ra$ are resonances and also nucleon-multi-pion states.   Each of these states has a detailed underlying structure in terms of quarks and gluons.
In  exclusive reactions  with not very large momentum transfer  few states are excited and  one may use \eq{sup} to describe the physics.  However, for high energy inclusive reactions of experimental relevance one needs many states. In this case a quark description is necessary.

\section{EMC effect-  discovery  at the small nuclear distance scale}

The aim of this Section is to exemplify  the connection between the small-distance scale related to virtuality and deep inelastic scattering from nuclei.\ The relation between virtuality and the EMC effect has been explored previously  in Refs.~\cite{Melnitchouk:1993nk,Kulagin:2004ie,CiofidegliAtti:2007ork,Hen:2016kwk,Segarra:2020plg}.\\

Deep inelastic scattering   (DIS) on a  free nucleon target was initially  expected to observe a set of 
resonances and therefore small cross sections for large values of three-momentum transfer ~\cite{Bloom:1969kc,Friedman:1972sy}.
Instead, the cross sections were large an approximate Bjorken scaling was observed. The unambiguous interpretation is that the nucleon contains quarks.\\

I explain in more detail. For typical DIS  kinematics 
 $Q^2=100 $ GeV$^2,\,x=0.5, \,\n\approx $100 GeV, the expansion  of \eq{sup} becomes unwieldy because the absorption of a virtual  photon by  free nucleon leads to a system of mass $M_X$ with $M_X^2=Q^2( {1\over x}-1) $, so $M_X\approx 10$ GeV. This high excitation energy tells us that  a huge number of  baryon states  are involved. Instead it is far more efficient to analyze the cross sections  using  quark degrees of freedom.   Measurements determine the quark structure functions  $q(x)$ that are scale and scheme dependent~\cite{Tanabashi:2018oca}. However, they are  well understood and interpreted as  momentum distributions.  Observe again that measurements of 
  experimental cross sections determine features of wave functions.\\

Next turn to deep inelastic scattering on nuclei at similarly large values of 
$Q^2$. It was initially  thought that at such kinematics  only very small  distances in the target would be involved~\cite{Aubert:1983xm}. Such distances are much, much less than the internucleon
spacing of $\approx$1.7 fm, and  the expectation  was  that  using a  nuclear target would only increase the number of target nucleons. 
Instead, the  medium modification of $q(x)$ was observed.  At high values of $x$ the ratio of the bound to free  structure function ratio is less than one by an amount of only between 10 and 15\%, dependent on the nucleus. This effect  is known as the EMC effect~\cite{Aubert:1983xm,Gomez:1993ri}. \\

That bound structure functions are different than free ones  is natural in terms of the discussion above regarding virtuality and \eq{sup}. Bound nucleons are virtual and the states $|N_n\ra$ have different structure functions than the nucleon.\\

Because of the large number of states entering in \eq{sup} it is most efficient to use quark degrees of freedom to understand DIS large values of   $Q^2$.
Then the   free nucleon is regarded as a
superposition of various configurations or Fock states, each with a different
quark-gluon structure.  \\

I  simplify the discussion  using a model  inspired by the QCD  physics of color transparency \cite{Frankfurt:1985cv,Brodsky:1987xw,Ralston:1988rb,Jennings:1993hw}.
The  infinite number of quark-gluon configurations of the proton are treated  as two configurations, a large-sized, blob-like  configuration, BLC, consisting of complicated configurations of many quarks and gluons, and a small-sized, point-like configuration, PLC, consisting of 3 quarks.
The BLC can be thought of as an object that is similar to a nucleon.
The PLC is meant to represent a  three-quark system of small size that is responsible for the high-$x$ behavior of
the distribution function. The smaller the number of  quarks, the more likely one can carry a large momentum fraction. 
 The small-sized configuration (with its small number
of $q\bar{q}$ pairs) is very different than a low lying nucleon excitation. This  two-component  model is meant to serve as a simple schematic tool to  enable qualitative understanding.  \\

When placed in a nucleus, the blob-like configuration feels the
usual  nuclear attraction and its energy decreases. The
point-like-configuration feels far less nuclear-attraction   by virtue of color screening ~\cite{Frankfurt:1994hf} in which the
effects of gluons emitted by small-sized configurations are cancelled
in low-momentum transfer processes.     The nuclear
attraction increases the energy difference between the BLCs and the
PLCs, therefore reducing the PLC
probability~\cite{Frankfurt:1985cv}. Reducing the probability of PLCs in the nucleus reduces the
quark momenta, in qualitative agreement with the EMC effect.\\

Working  out the consequences of the BLC-PLC model  enables the connection between the EMC effect and virtuality to be clarified.
  The
 Hamiltonian  for a free nucleon  in the two-component model can be expressed schematically by the matrix 
\bea H_0=
\left[ \begin{array}{cc}
    E_B & V \\
    V & E_P \end{array} \right], \eea 
where $B$ represents BLC and $P$ the PLC.  The  PLC is spatially much smaller than the BLC, so that   $E_P\gg E_B$. 
The hard-interaction potential, $V$, connects the two
components, causing   the eigenstates of $H_0$ to be  $|N\ra$ and $|X\ra$  rather
than $|B\ra$ and $|P\ra$.  In
lowest-order perturbation theory, the 
eigenstates are given by 
\bea 
|N\ra&=&|B\ra  +\e|P\ra, \label{nuc0}\\
|X\ra&=&-\e|B\ra +|P\ra,  
\eea
with
$
\e=V/(E_B-E_P)\ll1. \label{eps} 
$ 
It is natural to assume $|V|\ll E_P- E_B$, so that  the nucleon is mainly $|B\ra$ and its excited
state is mainly $|P\ra$. The notation $|X\ra$  is used to denote the state that is mainly a PLC, which  does not at all resemble a low-lying baryon resonance.\\

The quark structure function is  the matrix element of the operator $\cal O_{\rm DIS}$ that is the imaginary part of the virtual-photon- quark Compton scattering amplitude. This operator   acts on a single quark, so that 
\bea q(x) = {1\over 1+\e^2}\left(\la B|{\cal O}_{\rm DIS}|B\ra +\e^2 \la P|\CO|P\ra\right),\eea 
in which it is assumed that the single-quark operator 
 does not connect the two very different states $|B\ra$ and $|P\ra$.  Furthermore, the condition that the PLC dominates the structure function at large values of $x$ is enforced by defining a function $f(x)>0$ that monotonically increases as $x$ increases. In particular,
 let
 \bea
  \la P|\CO|P\ra\equiv f(x) \la B|\CO|B\ra,\label{fdef}\eea
  so that
  \bea q(x)=  {1\over 1+\e^2}\la B|\CO|B\ra (1+\e^2f(x)).\label{qdef}\eea
   \\

The model quark distributions of \cite{deTeramond:2018ecg}, based on light-front holographic QCD,
may provide a realization of the simple relation \eq{fdef}. These  incorporate Regge behavior at small $x$ and inclusive counting rules as $x$ approaches unity and is consistent with DIS  measurements.
  The model  provides    
quark distributions  $q_\tau(x)$ (normalized to unity)  as function of $\t$, the number of constituents in the system:
\bea  &
q_\tau(x) =\frac{\Gamma \left(\tau -\frac{1}{2}\right)}{\sqrt{\pi } \Gamma (\tau -1)}\big(1- w(x)\big)^{\tau-2}\, w(x)^{- \half}\, w'(x), \label{qt}
  \eea
with $w(x) = x^{1-x} e^{-a (1-x)^2}.$  The elastic form factors of this model  fall asymptotically as $1/Q^{2\t}$,  and the slope of form factors as $Q^2=0$ is proportional to $\t$. These features  mean  that an increase in the value of $\t$ corresponds to an increase in  effective size.
The function $q_3$ represents  a three quark system and is naturally associated with the  PLC.
\\

In \eq{qt}  the function $q_\t$ is normalized to unity.
The $u$ and $d$ quark distributions at a scale $\m_0=1.06\pm 0.15 $ GeV   are given by
\bea &
u(x) ={3\over 2}q_3(x) +{1\over 2} q_4(x)\\
&d(x)=q_4(x)
,\eea
with the $u(x)$ and $d(x)$ normalized to the flavor content of the proton.  An excellent reproduction of measured structure functions and elastic form factors is obtained using only two components and the   flavor-independent parameter  $a = 0.531 \pm 0.037$. This gives  some justification to the simple two-state picture of the present model.\\

The ratio $q_3(x)/q_4(x)=1/(1-w(x))$ which increases monotonically with increasing $x$, as expected by the intuition inherent in \eq{fdef} with $df/dx>1$. It is therefore reasonable to associate the PLC ($q_3$) with becoming more important as the value of $x$ increases. In this model  BLC is associated  with $q_4$, and 
the PLC component occurs only with up quarks. 
The relevant combination for a nucleus with $N$ neutrons and $Z$  protons is proportional to $Z{3\over2} (q_3(x)+ q_4(x))+{N\over4}(3q_3(x)+9q_4(x))$. \\

 Now suppose the nucleon is bound to a nucleus. The nucleon feels an attractive nuclear potential, here represented by  $H_1$, with 
 \bea H_1= \left[ \begin{array}{cc}
U & 0 \\
0 & 0  \end{array} \right]
 ,\eea
to represent the idea that  only the large-sized component of the nucleon feels the influence of the nuclear attraction. The treatment of the nuclear interaction, $U$,  as a number is clearly a simplification because 
the interaction necessarily varies with the relevant kinematics. The present  model is similar to 
   the model of~\cite{Frankfurt:1985cv}, with the important difference that the medium effects enter as an amplitude instead of as a probability.
See also Ref.~\cite{Frank:1995pv}.\\

The complete Hamiltonian $H=H_0+H_1$ is:
\bea H= \left[ \begin{array}{cc}
E_B-|U| & V \\
V & E_P  \end{array} \right],
 \eea
in which the attractive nature of the nuclear  binding potential is
emphasized. Then interactions with the nucleus increase the energy difference between
the bare BLC and  PLC states and thereby   decreases the PLC
probability.  \\

The medium-modified nucleon and its excited state,
$|N\ra_M$ and $|X\ra_M$, are now (again using first-order perturbation theory)
\bea 
|N\ra_M &=& |B\ra +\e_M|P\ra\ \label{mod1}\\
|X\ra_M&=& -\e_M|B\ra +|P\ra, 
\eea
where
\bea
\e_M={V\over{E_B-|U|-E_P}}= \e{E_B-E_P\over E_B-|U|-E_P}
\eea
and   ${\e_M\over\e}={E_B-E_P\over E_B-|U|-E_P} <1$.
 \\

The difference 
\bea
\e_M-\e\approx {|U|\over E_B-E_P}
\label{epsm}
\eea
is relevant for understanding the EMC effect
because
\bea |N\ra_M=|N\ra+(\e_M-\e)\la P|\CO|P\ra
,\eea
and the medium modification of the nucleon is proportional to  the interaction with the nucleus represented by $U$.\\

The medium-modified  quark distribution function $q_M(x)=\la N_M|{\CO}|N_M\ra$,
and is $q_M(x)=q(x)+\D q(x)$ with
\bea \D q\approx  2(\e_M-\e) \la N|{\CO}|P\ra\nonumber\\
 \approx  2(\e_M-\e) \e  \la P|{\CO}|P\ra
.\label{dq}\eea
in which terms of first-order in $(\e_M-\e)$ kept to represent the small EMC effect.
Next use \eq{fdef} and \eq{qdef} to find
\bea&
\D q(x)={2(\e_M-\e)\e }\,{q(x)f(x)\over 1+\e^2f(x)}\nonumber\\&
\approx {2(\e_M-\e) \e}\, q(x)f(x)
\eea
Note that the product $(\e_M-\e) \e$ is less than zero, independent of the sign of the interaction $V$. This means that, at large values of $x$,  the quark structure function in the  nucleus is less than that of a free nucleon, and decreases with increasing $x$ because $f(x)$ is monotonically increasing with increasing $x$. These features are inherent in the data for values of $x<0.7$. \\

The next step is to relate $(\e_M-\e)\propto U$ (via \eq{epsm})  to the virtuality. 
 Suppose a photon interacts with a virtual nucleon of four-momentum $ \bfP_{\rm miss}$
 The three-momentum
 $\bfP_{\rm miss}$
opposes the $A-1$ recoil momentum $\bfp\equiv \bfP_{\rm miss}=-\bfP_{A-1}$. The  mass of the on-shell recoiling nucleus is given by
  $M_{A-1}^*=M_A-M+E,$ where $E>0$ represents the excitation energy of
  the spectator $A-1$ nucleus, to  find~\cite{CiofidegliAtti:2007ork}
  \begin{eqnarray}
 &M^2 \Cv=
 \Pm^2-M^2\\&
 =(M_A-\sqrt{(M_{A-1}^*)^2 +\bfp ^2}\,)^2
- \bfp^2-M^2
 \end{eqnarray}
which reduces  in the non-relativistic limit to
  \begin{eqnarray}
M^2 \Cv&\approx& -2M\left({\bfp^2\over 2M_r}+E\right),
\label{virt}
\end{eqnarray}
where the reduced mass  $M_r= M
(A-1)/A$. The virtuality,  $\Cv$,  is less than 0, and its 
 magnitude
 increases with both the $A-1$ excitation energy and the  initial momentum of the struck nucleon.\\

Refs.~\cite{Frankfurt:1985cv,CiofidegliAtti:2007ork} obtained a relation between
the potential $U$ and the virtuality
$\Cv$ by using the extension of the Schroedinger equation to an
operator form: 
\bea {\bfp^2\over 2M_r}+U=-E,
\eea 
so that $ {\bfp^2\over 2M_r}+E=-U=|U|$ and via \eq{epsm}
 \bea \Cv = {2U\over M}=
 {2(\e_M-\e)(E_P-E_B)\over M} ,\label{cv}
\eea
so that    the modification of the nucleon due to the PLC suppression is proportional to its virtuality. Potentially large values of the virtuality  greatly enhance the difference between  $\e_m$ and $\e$. \\

Recall \eq{dq} and replace $(\e_M-\e) $  therein by its expression in terms of $\Cv$ (\eq{cv}) to find
\bea  q_M(x)= q(x) +{M\over E_P-E_B}\Cv \e\,f(x) q(x),
\eea
The conditions that $(\e_M-\e) \e<0,\,\Cv<0$ and \eq{cv} lead to the requirement that $\e>0$, which means that $V<0$. The sign of $\e$ is consistent with the light-front holographic model for which $\e=1/\sqrt{2}$ for the proton and 0 for the neutron. 
The suppression of point-like components is  manifest by the condition $ df/dx>0$ and $\e df/dx>0$.
The ratio of structure functions is   $R(x)=q_M(x)/q(x)$, and
\bea 
{dR\over dx}&=&{M\over E_P-E_B}\Cv\e{d f\over dx}<0, \label{fxx}
\eea
as  the measurements of the EMC effect have shown. The negative sign is caused by the negative value of the virtuality.
This expression is only meaningful for $x<0.7$ where Fermi motion effects can be ignored.
\\

\begin{widetext}
\begin{center}
\begin{table}[h!]
    \caption{EMC effect {\it vs.} Virtuality.} 
    \label{tab:table1}
    \begin{tabular}{lccccc} 
      {Quantity  } & {$^3$ He  } & {$^4$He }& $^{12}$C & $^{56}$Fe& $^{208}$Pb\\
        \hline
     $|{dR\over d x}| $  \cite{Weinstein:2010rt}. & 0.070 $\pm$0.029 & 0.197$ \pm0.026$&0.292$\pm$0.023& 0.388 $\pm$0.032&0.409$\pm$ 0.039\\
     $|{\Cv\over 2M}|$ (MeV) \cite{CiofidegliAtti:2007ork} & 34.59 & 69.4& 82.28&82.44 &92.2\\
 \hline
    \end{tabular}
\end{table}
  \end{center}

\end{widetext}

The quantities $M, E_P-E_B$ and $f(x)$ are independent of the nucleus, so that the $A$-dependence of the EMC effect is determined by the virtuality, $\Cv$.
According to this model, the larger the virtuality the larger the EMC effect, as measured by the slope of $R(x)$.
Table I compares the measurements of the slope with computations of the virtuality.  The data for A=56 is from a mixture of A=56 and A=63. The theory for $^{208}$Pb is compared with the data for $^{197}$Au.   The increase of the magnitude of the slope tracks qualitatively well with the corresponding increase of the virtuality. 
A   quantitative reproduction of the A-dependence requires a more detailed treatment of the separate N and Z dependence as in  Ref.~\cite{Schmookler:2019nvf}.\\

  Another consequence of this  model  is that the  medium-modified nucleon contains a component that  is an excited state of  a free nucleon. The amount of modification, $\e_M-\e$, which gives a  deviation of the EMC ratio from
unity, is controlled by   the
  potential $U$ and via \eq{cv} the virtuality. A more detailed evaluation of the EMC effect is reserved for another paper.
  \\
  


\section{Summary \& Discussion}
This paper takes a trip through three length scales relevant to nuclear physics. These are the nuclear size, the inter-nucleon separation distance and the nucleon size. Simple examples are used to illustrate the basic underlying features that drive the observations made at the three different scales.
The intent is to arrive at the realization that all three scales are must be understood to truly understand the physics of nuclei. \\

Sect.~II briefly reviews the currently popular procedure of softening the interactions between nucleons, with a focus on the concomitant hardening of the operators that probe nuclei.
A first-order equation, \eq{fo} is derived to demonstrate that the probe operators are  hardened by the same unitary transformation that softens the interactions.\\

Sec.~III discusses the largest nuclear scale, with the first point being that momentum transfers higher than that achieved by Rutherford were needed to discern the non-zero nature of the nuclear size. Equations  (\eq{foA} and \eq{foa1}) are  derived to estimate the effect of the hardening of the probe operator, and is used to demonstrate its importance for momentum transfers, $q$, greater than about 2 fm$^{-1}$.\\

The physics of the nucleon-nucleon separation is explored in Sect.~IV by using bound-state wave functions produced by  four simple models of the nucleon interaction.
The high-momentum transfer ($q$) scaling of  the form factors  is exhibited for each model.  The values of relative momentum $p$ that make important contributions to the form factor are displayed.  Increasing the value of  $q$ is shown to increase the values  of $p$ that enter. The resulting  effect of the hardening of the probe operator is displayed for two of the model interactions, where again significant effects of hardening of the operator are seen for $q>2$ fm$^{-1}$. For other interactions the hardening cannot be computed easily. The role of the tensor force in producing high-momentum components, and in transforming the probe operator,  is also discussed.  Current experiments involve transfer of high momentum. The interpretation of such experiments is simplified if bare, un-transformed probe operators can be used. \\

The role of two-nucleon physics in nuclei, as manifest in the independent pair approximation,  is explored in Sect.~V.  The modern approach is the generalized contact formalism. The high-momentum properties of  0-energy wave functions entering that formalism are examined. The result \eq{ae} demonstrates the explicit connection between short-distance and high-momentum physics. Furthermore, the inevitable power-law falloff indicates that significant high momentum content must occur. The conditions necessary for obtaining a direct connection, \eq{wow}, between scaling behavior of  measured form factors and the underlying wave functions are determined.  \\

Sect. VI discusses the $(e,e',p)$ reaction as a tool for discovery of short-distance physics at the nucleon-nucleon separation scale. Under certain conditions \eq{imp}, which directly relates the scattering amplitude to the wave function, is valid. More generally, at high momentum transfer, final state nucleons have high energy and undergo different interactions than those in the initial state. Thus, in such situations, it is far simpler to use the impulse approximation with the fundamental potentials in the Hamiltonian than to use interactions softened by unitary transformations. \\

The transition from the nucleon-nucleon separation distance  to the nucleon size and  smaller sizes is begun in Sect.~VII through a discussion of virtuality, \eq{v0}.  High momentum transfer reactions probe highly virtual nucleons. 
Nucleons achieve high virtuality only through strong interactions with closely separated nucleons, \eq{small}.
The internal wave function of such nucleons may be expressed as a superposition of baryonic eigenstates, \eq{sup}. If the momentum transfer is large enough many, many  states must be included in the superposition, and it becomes more efficient to use quark degrees of freedom.\\

The role of virtuality in understanding the nuclear modification of quark structure functions (EMC effect)  is discussed in Sect.~VIII. The explicit connection, \eq{fxx}  is displayed by using  a two-component, (point-like/blob-like) model of the nucleon's quark degrees of freedom. 
The simple model is shown to be consistent with the two-state treatment of light-front holographic QCD that reproduces free nucleon structure functions and elastic form factors. In particular, the point-like component is more important relative to the blob-like component at larger values of $x$. This model, combined with the concept of virtuality provides a qualitative explanation of the EMC effect. 
\\

\section*{Acknowledgements}

 This work was supported by the U. S. Department of Energy Office of Science, Office of Nuclear Physics under Award Number DE-FG02-97ER-41014.  I thank S.~R.~Stroberg  and X-D Ji for  useful discussions.
\section{Appendix-Derivation of \eq{fo}}
The result, \eq{fo} is stated without treating  the term of first order in $s$ caused by the $s$-dependence of the potential. This Appendix shows that the term  vanishes for the case of a local, bare potential and a local operator $\Oop$.\\

Consider the matrix element 
\bea {\cal M}_s\equiv \la\Psi|[H_0,[V_s,\Oop]]|\Psi\ra,\eea
which enters in computing elastic form factors.
The goal here is to show that the term of order $s$ vanishes.
To first order in $s$ 
\bea
V_s\approx V+s{dV\over ds}(s=0)= V+ s [[H_0,V],V].
\eea
The double commutator $[[H_0,V],V]= {-2\over M}(\bfnabla V)^2$ which is function of $\bfr$. This commutes with $\Oop$ and  \eq{fo} is obtained.  
 
\bibliography{miller}{}

\end{document}